\documentclass[twocolumn,superscriptaddress,pra,preprintnumbers,showpacs,amsmath,amssymb]{revtex4}

\usepackage{graphicx}

\newcommand{\la}{\langle}
\newcommand{\ra}{\rangle}
\newcommand{\diff}{\mathrm{d}}
\newcommand{\sinc}{\text{\rm sinc}}
\newcommand{\vp}{\boldsymbol{p}}

\newcommand{\vx}{\boldsymbol{r}}
\newcommand{\ex}{\boldsymbol{e}_x}

\newcommand{\ez}{\boldsymbol{e}_z}

\newcommand{\oU}{ \mathsf U }
\newcommand{\oS}{ \mathsf S }

\newcommand{\oP}{ \text{\sf \textbf{p}} }

\newcommand{\U}{ {\cal U} }
\newcommand{\V}{ {\cal V} }
\newcommand{\Order}{ {\cal O} }

\begin{document}

\title{Theory of near-field matter wave interference beyond the eikonal approximation}

\author{Stefan Nimmrichter}
\affiliation{%
Arnold Sommerfeld Center for Theoretical Physics,
Ludwig-Maximilians-Universit\"{a}t M\"{u}nchen,\\
Theresienstra{\ss}e 37, 80333 Munich, Germany
}
\affiliation{Faculty of Physics, University of Vienna, Boltzmanngasse 5, 1090 Vienna, Austria}

\author{Klaus Hornberger}%
\homepage{www.klaus-hornberger.de}
\affiliation{%
Arnold Sommerfeld Center for Theoretical Physics,
Ludwig-Maximilians-Universit\"{a}t M\"{u}nchen,\\
Theresienstra{\ss}e 37, 80333 Munich, Germany
}%

\preprint{\sf published in: Phys.~Rev.~A \textbf{78}, 023612 (2008)}

\begin{abstract}
A generalized description of Talbot-Lau interference with matter waves is presented, which accounts for arbitrary grating interactions and realistic beam characteristics. 
The dispersion interaction between the beam particles and the optical elements strongly influences the interference pattern in this near-field effect, and it is known to dominate the fringe visibility if increasingly massive and complex particles are used.
We provide a general description of the grating interaction process by combining semiclassical scattering theory with a phase space formulation.  It serves to systematically improve the eikonal approximation used so far, and to assess its regime of validity.
\end{abstract}


\maketitle

\section{Introduction}

The ability of material particles to show wave-like 
interference is one of the central predictions of quantum mechanics. While the early experimental tests worked with elementary particles \cite{Davisson1927a,Halban1936a}, interferometry of atoms is by
now a matured field of physics \cite{Berman1997a,Miffre2006a,Cronin2008a}.
It is in particular the ability to cool and to control atoms using laser techniques
that has propelled atom interferometry into a versatile tool for
precision measurements, e.g. \cite{Weiss1993a,Gustavson1997a,Peters1999a,Muller2007a}.

As far as more complex and more massive objects are concerned, we are
currently witnessing this transition from the proof-of-principle
demonstration of their wave nature to the use of interferometry for
quantitative measurements.  Specifically, the static
\cite{Berninger2007a} and the dynamic \cite{Hackermuller2007a} bulk
polarizability of fullerene molecules was measured recently with a
molecule interferometer. Also the controlled observation of
decoherence, due to collisions with gas particles
\cite{colldecoboth} or due to the emission of
thermal radiation \cite{thermodecoboth}, has been
used to characterize the interaction strength (or cross-section)
of fullerenes with external degrees of freedom. Another motivation for
studying interference with large molecules is to test quantum
mechanics in unprecedented regimes by establishing the wave nature of ever more massive objects \cite{Arndt2005a}.

The above-mentioned interference experiments with large molecules
are based on the near-field Talbot-Lau effect, where three
gratings are used which serve, in turn, to produce coherence in
the beam, to bring it to interference, and to resolve the fringe
pattern.  This is an established technique in atom and electron
interferometry
\cite{Clauser1994both,Cahn1997a,Deng1999a,Cronin2006a,Kohno2007a}, and it is the
method of choice for massive and bulky molecules (such as fullerenes
\cite{Brezger2002a}, meso-tetraphenylprorphyrins \cite{Hackermuller2003a},
or functionalized azobenzenes \cite{Gerlich2007a}). The main reason is that a
Talbot-Lau interferometer (TLI) tolerates beams which are relatively weakly
collimated, thus alleviating the increasing difficulty in producing
brilliant beams if the particles get more complex.
Another important advantage compared to far-field setups, which is
essential if one wants to increase the particle mass by several orders
of magnitude, is the favorable scaling behavior of Talbot-Lau
interferometers with respect to the de Broglie wavelength \cite{Berman1997a}.

As a specific feature of Talbot-Lau interferometers, the forces between the particles in the beam and the diffraction grating
influence the interference pattern much stronger than in a far field
setup. This is due to the fact that the different diffraction orders
do not get spatially separated in a TLI. Rather, all the orders
interfere among each other, producing a resonant recurrence of the
pattern whenever the so-called Talbot condition is met. As a
consequence, even tiny distortions of the matter wave may lead to
significant changes of the fringe visibility---requiring, for example,
the modification of the van der Waals force due to retardation effects
\cite{Casimir1948a} to be taken into account when describing the
diffraction of fullerenes at gold gratings.

This effect of the dispersion force between the particle and the
grating wall gets more important as the mass and structure of the
molecule grows. In particular, it sets increasingly strict
requirements on the monochromaticity, i.e, the velocity spread permissible in the
molecular beam.  A very recent development, undertaken to reduce this
influence, is the Kapitza-Dirac Talbot-Lau Interferometer (KDTLI),
where the second material grating is replaced by the pure phase grating
produced by a standing light field \cite{Gerlich2007a}.

The available theoretical descriptions of molecular Talbot-Lau
interference account for the grating forces in terms of a simple
eikonal phase shift \cite{Patorski1989a,Brezger2003a,Hornberger2004a}
(an expression originally derived by nuclear physicists as an
asymptotic high-energy approximation to the Lippmann-Schwinger
equation in scattering theory \cite{Moliere1947a,Glauber1959a}).
This approximation ceases to be valid with a growing influence of the
particle-grating interaction, and, due to its non-perturbative nature,
its range of validity is not easy to assess. Therefore, given the quest
for testing quantum mechanics with ever larger particles and given the
increased precision required in metrological applications, there is a
clear need to extend the theoretical description of near-field matter
wave interference beyond the eikonal approximation.

The main purpose of this paper is therefore to develop a generalized
formulation of the coherent Talbot-Lau effect. By combining a
scattering theory formulation with semiclassical approximations we
incorporate the effect of the grating interaction systematically 
beyond the eikonal approximation.
As a prerequisite for its implementation, the established theory of near
field interference first needs to be extended to account for the
effects of finite angular dispersion in the molecular beam. We show
how this can be done transparently by using the phase space formulation
of quantum mechanics. As a by-product, this formulation permits us to
quantify the adjustment precision required in realistic experiments.

The structure of the article is as follows. In Section
\ref{sec:phasespace} we develop a generalized theory of Talbot-Lau
interference, which is formulated independently of the particular
choice of how to incorporate the grating interaction. We also establish the relation to
the previous treatments by applying the
eikonal approximation. In Section \ref{sec:potential},
we review relevant realistic descriptions of the grating interaction, 
and numerically illustrate their effect in the eikonal approximation.
Section \ref{sec:semiclass} is devoted to the development of a
semiclassical formalism to go beyond the elementary eikonal approximation.
Its predictions are numerically evaluated and compared in Section
\ref{sec:numerics}, using the experimental parameters of the
molecular interference experiments carried out in Vienna
\cite{Brezger2002a,Gerlich2007a}. Finally, we present our conclusions in
Section \ref{sec:conclusions}.

\section{Phase space description of the Talbot-Lau effect} \label{sec:phasespace}

The effect of Talbot-Lau interference can be described in a
particularly transparent and accessible fashion by using the phase
space representation of quantum mechanics
\cite{Wigner1932a,Ozorio1998a,Schleich2001a,Zachos2005a}. This is demonstrated in
\cite{Hornberger2004a}, where both the coherent effect and the
consequences of environmental interactions are formulated in terms of
the Wigner function of the matter wave beam. As also shown there, the
analogy of the Wigner function with the classical phase space
distribution allows one to evaluate the predictions of classical and
quantum mechanics in the same framework, a necessary step if one wants
to distinguish unambiguously quantum interference from a possible
classical shadow effect.

However, the treatment in \cite{Hornberger2004a} is based on a number
of idealizations, which must be reconsidered in view of a more refined
description of the particle-grating interaction. The least problematic
approximation is to disregard the motion in the 
direction parallel to the grating slits, which is permissible due
to the translational symmetry of the setup in this direction. It
follows that a two-dimensional description involving the
longitudinal motion (denoted by $z$) and the transverse motion (denoted by $x$) of the beam is required in principle. In front of the interferometer these two degrees of freedom
are well approximated by a separable state involving transverse momenta $|p|\ll p_z$. 
If the grating interaction is treated in eikonal approximation,
as done in \cite{Hornberger2004a}, this implies that the transverse
and the longitudinal motion remain separable throughout. One can then
resort to an effectively one-dimensional description, characterized by
a fixed longitudinal momentum $p_z= h /\lambda$. The $z$ coordinate
then represents a time $t=m z /p_z$ and the longitudinal propagation
of the beam along $z$ effectively evolves the one-dimensional
transverse beam state during the time $t$.  The finite distribution
$\mu(p_z)$ of the longitudinal momenta is then accounted for only in
the end by averaging the results obtained with sharp values of $p_z$.

A priori, such a treatment is no longer valid for a general grating
interaction where different longitudinal momentum components of the
beam get correlated. We will accordingly use a two-dimensional
scattering formulation to describe the grating interaction in
Sect.~\ref{sec:semiclass}. However, we will see that for the
parameters of typical experiments the main effect of taking the
grating interaction beyond the eikonal approximation is on the
transverse degrees of freedom. Since the effect on the longitudinal
motion is much weaker we will retain the effectively one-dimensional
description outlined above, postponing the physical discussion why
this is permissible to Sect.~\ref{sec:semiclass}.

Another idealization found in the basic treatments of the Talbot-Lau
effect is to assume the transverse motion in front of the interferometer to be in a completely incoherent state. This would correspond to a constant distribution of the transverse momenta $p$, and for general grating transformations, which depend on $p$, it is no longer a valid approximation. We will therefore present a formulation that takes into account a realistic beam profile and that allows for a general state transformation at the diffraction grating. Going beyond the idealization of a perfectly incoherent state will also permit us to assess the adjustment precisions required in an experimental implementation.

\subsection{The Wigner function and its transformations}

We start by briefly outlining how to describe matter wave interference by means of the phase space representation of quantum mechanics \cite{Hornberger2004a,Hornberger2006a}. As discussed above, one may restrict the dynamics to the transverse beam state, which is most generally specified by its density matrix $\rho$. This state is equivalently described by the Wigner function
\begin{equation}
 w \left(x,p \right) = \frac{1}{2 \pi \hbar} \int \diff s \, e^{i p s / \hbar} \la x - \frac{s}{2}|\rho|x + \frac{s}{2} \ra,
\end{equation} 
which is a function of the transverse phase space coordinates $\left(x,p \right)$. 
Like $\rho$ it depends parametrically on the longitudinal momentum $p_z$. Since we assume the beam to be collimated $w(x,p)$ is non-zero only for $|p|\ll p_z$.

The great advantage of the phase space formulation is that it permits a straightforward, yet realistic, description of the beam and its propagation through
the interferometer. Most importantly, the free time evolution of a state
during the time $t$ is given by the same shearing transformation 
as in the case of the classical phase space density,%
\begin{equation}
w_t \left(x,p \right) = w_0 \left( x-\frac{t}{m} p ,p \right),
\label{eq:wt}
\end{equation} 
with $m$ the particle mass. 
On the other hand, a quantum state transformation of the form $\rho' = \oU \rho
\oU^{\dagger}$, such as the effect of passing a grating, reads in phase space 
\begin{equation}
 w' \left(x,p \right) = \iint \diff x_0 \diff p_0 \, K \left(x,p;x_0,p_0 \right) w \left(x_0,p_0 \right).
\end{equation} 
Here the integral kernel is given by the the propagator 
\begin{eqnarray}
 K \left( x, p; x_0 , p_0 \right) &=& \frac{1}{2\pi \hbar} \iint \diff s \diff s_0 \, e^{i \left(p s + p_0 s_0 \right)/\hbar} \nonumber \\
&&\times \la x - \frac{s}{2} |\oU| x_0 + \frac{s_0}{2} \ra \la x + \frac{s}{2} |\oU| x_0 - \frac{s_0}{2} \ra^{*} .
 \nonumber \\ &&
\label{eqn:generalprop}
\end{eqnarray} 
Specifically, an ideal
grating is characterized by a grating transmission function $t(x)$, with $|t(x)|\leq 1$, describing the multiplicative modification of an incoming plane wave. (The factor $t(x)$ is non-zero only within the slit openings of the grating and it may there imprint a complex phase to account for the interaction potential between the grating walls and
the beam particle, see Sect.~\ref{sec:potential}.) For such gratings the transformation reads  $\la x |\oU |x_0 \ra = t \left(x \right) \delta\left(x-x_0 \right)$, so that the grating propagator reduces to a convolution kernel 
\begin{eqnarray}
 K \left( x, p; x_0 , p_0 \right) &=& \delta \left(x-x_0 \right) \frac{1}{2\pi \hbar} \int \diff s \, e^{i \left(p - p_0\right) s/\hbar} \nonumber \\
&&\times t \left(x- \frac{s}{2} \right) t^{*} \left(x+ \frac{s}{2} \right) ,
\label{eqn:eikonalkernel}
\end{eqnarray}
which is local in position \cite{Hornberger2004a}. This choice of $K$ will be required below to reduce the generalized formulation of Talbot-Lau interference to the eikonal approximation. 

The convolution kernel provides a descriptive picture
of the diffraction process. Suppose the incoming beam is perfectly coherent, i.e. a plane wave characterized by the longitudinal momentum $p_z$ and transverse momentum $p_0$, corresponding to the (unnormalized) transverse Wigner function $w_0 \left(x,p\right) = \delta \left(p-p_0\right)$. The diffracted transverse state is then given by
\begin{equation}
  w_1 \left(x,p \right)	= \frac{1}{2\pi \hbar} \int \diff s \, e^{i \left(p - p_0\right) s/\hbar} t \left(x- \frac{s}{2} \right) t^{*} \left(x+ \frac{s}{2} \right).
\end{equation}
In case of a periodic grating with period $d$, the transmission function can be
decomposed into a Fourier series, %
\begin{equation}
t (x) = \sum_{j=-\infty}^\infty b_j \exp \left( 2\pi i j \frac{x}{d} \right),
\label{eq:bj}
\end{equation}
so that, after a free propagation over the longitudinal distance $L$, the 
transverse spatial density of the beam state is given by the marginal distribution
\begin{eqnarray}
	w_2 \left(x \right) &=& \int \diff p \, w_1 \left( x- \frac{p}{p_z}L,p\right) \nonumber \\
	&=& \sum_{m=-\infty}^{\infty} 
	B_m\left(m \frac{ L}{L_\textrm{T}}\right)
	\exp \left( 2\pi i m \frac{x}{d} \right).
\label{eq:TE}
\end{eqnarray}
Here, we introduced the basic \textit{Talbot-Lau coefficients},
\begin{equation}
	B_{m} \left(\xi \right) = \sum_{j=-\infty}^{\infty} \, b_j b_{j-m}^{*} \exp \left(i \pi \xi \left( m - 2j\right) \right),
	\label{eqn:Beik}
\end{equation}
and the characteristic length scale $L_{\rm T} = d^2 / \lambda$ is called the Talbot length. 

Compare (\ref{eqn:Beik}) to the Fourier coefficients of the transmission probability $|t(x)|^2$, which are given by the convolution $A_{m}=\sum_j b_j b_{j-m}^{*}$. Equation (\ref{eq:TE}) thus implies that the density distribution takes the form of the grating transmission profile whenever the distance $L$ is an integer multiple of the Talbot length, $L=k
L_{\textrm{T}}$,
\begin{equation}
	w_2 \left(x \right) = \left| t \left( x + k \frac{d}{2}\right)\right|^2.
	\label{eqn:Talboteffect}
\end{equation}
This is the elementary Talbot effect \cite{Talbot1836a}, and it is the backbone of the Talbot-Lau interferometer, which however does not require a coherent illumination.

\subsection{The general Talbot-Lau effect} \label{sec:generalTL}

\begin{figure*}
\includegraphics[width=15cm]{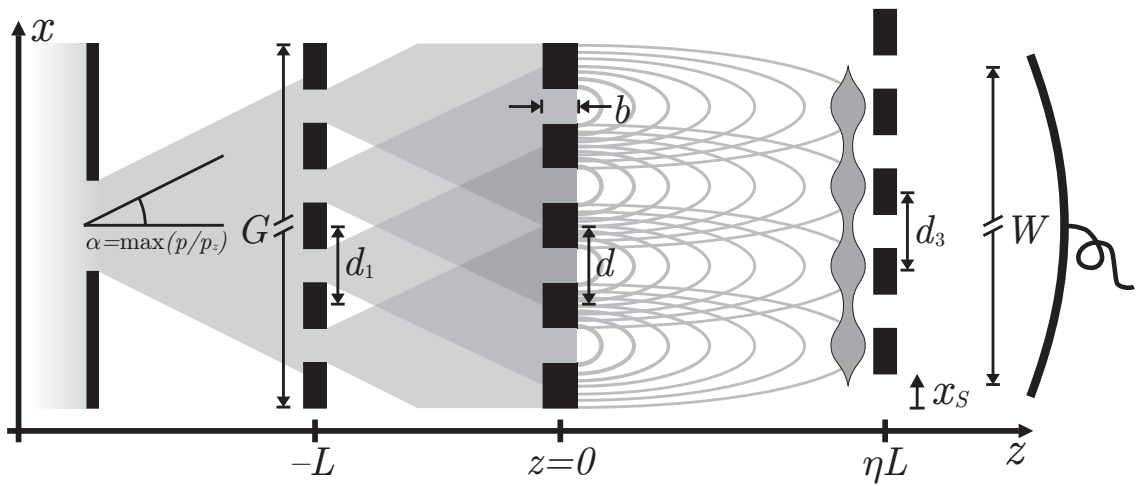}
\caption{\label{fig:TLsetup} Schematic of a Talbot-Lau interferometer with the relevant setup parameters for a material diffraction grating. The beam emanating from the left is constrained by collimation slits to an angular spread $\alpha$ before it enters the interferometer. The first grating of period $d_1$ and size $G\gg d_1$ modulates the beam so that spatial coherence is created at the position of the second grating ($z=0$) where the diffraction takes place. This leads to an interference pattern in the transverse beam density, ideally located at the position of the third grating. The latter can be shifted in the transverse direction by a variable distance $x_S$. It thus modulates the flux provided its longitudinal position $\eta L$ and its period $d_3$ match the interference pattern. A detector of size $W$ further downstream measures the total incoming beam intensity as a function of $x_S$. The central grating may be replaced by the phase grating created by a standing light field.}
\end{figure*}

The general setup of a Talbot-Lau interferometer is described in Fig.~\ref{fig:TLsetup}, along with the most relevant parameters. An incoherent particle beam represented by the shaded area
passes a preliminary collimation slit on the left side of the figure and enters the Talbot-Lau interferometer through the first grating. 
The angular distribution of those particles in the beam which are finally detected is characterized by the spread $\alpha$, 
determined by the collimation slit and the finite area of the detector.

The first grating can be understood as an array of collimation slits of period $d_1$ preparing, after a distance $L$, the transverse coherence required for diffraction at the second grating. In this region,
the distortion of a matter wave front due to the grating interaction may significantly affect the interference pattern, so that the grating thickness $b$ enters the calculation. (It is replaced by the laser waist $w_z$ in case of a light grating.)

The different diffraction orders still overlap further downstream if the distance is on the order of the Talbot length $L_{\textrm{T}} = d^2 / \lambda$, and they may thus interfere among each other producing a near field fringe pattern. This can be explained qualitatively from the elementary Talbot effect (\ref{eqn:Talboteffect}) by considering each of the slits in the first grating as independent point sources. The Talbot patterns due to the individual slits will add up constructively for appropriate choices of the grating periods $d_1$, $d$ and of the distance factor $\eta$, so that a distinct density pattern is created. It is verified without the need for a spatially resolving detector by superposing a 
third grating whose period $d_3$ equals that of the density pattern, and by measuring the total flux through it as a function of its transverse position $x_S$.

We will now present a quantitative formulation of the successive propagation of the beam through the interferometer. As discussed above, one can take the longitudinal motion of the beam particles to remain unaffected by the gratings. We may therefore assume the longitudinal momentum to have a definite value $p_z$ for the time being, so that the longitudinal position $z$ plays the role of a time coordinate $t= z m /p_z$ for the transverse motion.

\subsubsection{Sequential calculation}

The transverse state of the beam entering the interferometer is far from pure. It is confined in position by the orifice of the first grating, whose size $G$ typically covers thousands of grating slits. The momenta $p$ are characterized by the angular distribution $D \left( p/p_z \right)$ whose characteristic spread is denoted as $\alpha$.

In a typical experimental situation, $\alpha$ rarely exceeds $1$\,mrad corresponding to a fairly well-collimated beam. However, this still covers a range of transverse momenta that is by orders of magnitude larger than the grating momentum $h/d_1$ which separates the different diffraction orders. This explains why diffraction does not need to be taken into account at the first grating. The transverse Wigner function merely gets modulated by the grating profile, so that behind the first grating it reads
\begin{equation}
 w_1 \left(x,p \right) = \frac{1}{G p_z} \left| t_1 \left(x \right) \right|^2 D \left( \frac{p}{p_z}\right).
\end{equation} 
Here $t_1 \left(x \right)$ is the transmission function which is confined to the grating orifice, in principle.
However, since the size $G$ is typically larger than
the sensitive region $W$ of the final detector
one may equally take it to be an unconstricted periodic function.

The free propagation of the beam over the longitudinal distance $L$ yields $w_2(x,p)=w_1 (x-L p/p_z,p)$.
The effect of passing the second grating is in general described by an operator $\oU$ to be specified below. The corresponding phase space transformation is given by (\ref{eqn:generalprop}) so that, after a second free propagation over the distance $\eta L$, the transverse beam state in front of the third grating reads
\begin{eqnarray}
 w_3 \left(x,p \right) &=& \frac{1}{G p_z} \iint \diff x_0 \diff p_0 \, D \left( \frac{p_0}{p_z} \right) \, \left| t_1 \left( x_0\right)\right|^2 \nonumber \\
&&\times \, K \left( x - \frac{p}{p_z} \eta L, p; x_0+\frac{p_0}{p_z}L, p_0 \right) .
\label{eq:w3eins}
\end{eqnarray}
The corresponding spatial density distribution, given by
$ w_3 \left(x \right) = \int \diff p \, w_3 \left(x,p \right)$,
is now modulated by the third grating as a function of the grating shift $x_S$. The detection signal is thus obtained as
\begin{eqnarray}
 S \left(x_S \right) &=& \int_{-W/2}^{W/2} \diff x \, w_3 \left(x \right) \left| t_3 \left(x-x_S \right) \right|^2
\nonumber\\
&=& \sum_{n=-\infty}^\infty S_n \exp \left( 2\pi i n\frac{ x_S}{d_3}\right),
 \label{eqn:signal}
\end{eqnarray} 
where $\left| t_3 \left(x\right) \right|^2$ is the spatial transmission probability of the grating and $W$ the size of the detector. The latter typically covers hundreds of grating periods so that
one may disregard the finiteness of $W$ when evaluating the basic interference effect. However, as shown below, it does play a role when the experimental adjustment precision needs to be evaluated. 

The resulting interference pattern is most easily characterized by the contrast of the modulation signal,
conventionally defined as the ratio between the amplitude of the signal variation and its mean value.
However, since the experimental signal is usually very close to a sinusoidal curve, it is in practice more convenient and more robust to use the \emph{sinusoidal visibility} defined in terms of the first to the zeroth Fourier expansion coefficient of the $d_3$-periodic signal \cite{Brezger2003a},%
\begin{equation}
	\V = \left| \frac{2 S_1}{S_0} \right| .
	\label{eqn:Vexp}
\end{equation}
It can be easily obtained from noisy experimental data by fitting a sine curve and it coincides with the conventional definition in case of a sinusoidal signal.

\subsubsection{Decomposing the grating propagator}

We proceed to evaluate the interference pattern (\ref{eq:w3eins}) by noting a general property of the grating transformation $\oU$. The periodicity of the grating implies that the position representation $\la x|\oU|x_0 \ra$ is $d$-periodic with respect to the center position $(x+x_0)/2$.
This admits the series expansion
\begin{equation}
 \la x| \oU |x_0 \ra = \sum_{n=-\infty}^{\infty} \, \exp \left(i \pi  n\frac{x+x_0}{d} \right) \, \U_n \left( x- x_0\right)
\end{equation} 
where the transformation within a single slit is now characterized by the corresponding Fourier coefficients
\begin{equation}
 \U_n \left( \tilde{x} \right) = \frac{1}{d} \int_{-d/2}^{d/2} \diff \bar{x} \, \exp \left(-2 \pi i n\frac{ \bar{x}}{d} \right) \la \bar{x} + \frac{\tilde{x}}{2} |\oU| \bar{x} - \frac{\tilde{x}}{2} \ra .
 \label{eq:Undef}
\end{equation} 
It is now convenient to specify these transformation functions in terms of their \emph{shifted} Fourier transformation, which serves to generalize the grating coefficients introduced for the eikonal case in (\ref{eq:bj}).%
\begin{equation}
 b_n \left( {p} \right) = \int \diff {x} \, \exp \left(-\frac{i}{\hbar} \left({p} + \frac{\pi \hbar}{d} n \right) {x} \right) \, \U_n \left( {x}\right) 
 \label{eqn:gratingcoeff}
\end{equation} 
Indeed, for position-diagonal operators $\oU$, these functions drop their 
momentum-dependence and reduce to the coefficients in (\ref{eq:bj}).

The generalized grating coefficients (\ref{eqn:gratingcoeff}) can now be used to construct the generalized Talbot-Lau coefficients, which are the central quantities for describing the interference effect.
 \begin{eqnarray}
B_{m} \left(\xi; p, \nu \right) &=& \sum_{j=-\infty}^{\infty} \, b_j \left( p +\nu\frac{\pi \hbar}{d} \right) b_{j-m}^{*} \left( p - \nu\frac{\pi \hbar}{d} \right) \nonumber \\
&&\times \, \exp \left(i \pi \xi \left( m - 2j\right) \right) ,
\label{eqn:TLcoeff}
\end{eqnarray}
We note that at integer values of the argument $\xi$ the phase factor in (\ref{eqn:TLcoeff}) reduces to a constant sign, while $p$ has the meaning of an incident momentum and $\nu$ that of a scale factor, as will be seen below.

By using the generalized Talbot-Lau coefficients the density distribution at the third grating takes the explicit form
\begin{eqnarray}
 w_3 \left(x \right) &=& \frac{1}{G} \sum_{m,n = -\infty}^{\infty}\, A_{n} \, \exp\left(2 \pi i \left( m + n \frac{d}{d_1} \right)\frac{ x}{d} \right) \nonumber \\
&& \times \, \int \frac{\diff p}{p_z} \, \exp \left(-2 \pi i\frac{p L}{p_z d} \left( \eta m + (\eta + 1) n \frac{d}{d_1} \right) \right) \nonumber \\
&& \times \, D \left( \frac{p}{p_z} \right) B_{m} \left( \eta \left( m+ n \frac{d}{d_1} \right) \frac{L}{L_{\textrm{T}}}; p, n \frac{d}{d_1} \right) .\nonumber
\\ & & \label{eqn:interferencepattern1}
\end{eqnarray} 
This expression for the interference pattern is now further simplified by identifying the Talbot-Lau resonance condition.

\subsubsection{The resonance condition}

Only a part of the double summation in (\ref{eqn:interferencepattern1}) contributes appreciably to the interference pattern. This follows from the fact that the ratio $L/d$ of grating distance to grating period is a large number, typically on the order of $10^5$. As a result, the momentum dependence of the phase factor in (\ref{eqn:interferencepattern1}) occurs on a very different scale compared to the variation of the momentum distribution $D \left( {p}/{p_z} \right)$ and of the Talbot-Lau coefficients $B_m(\xi;p,\nu)$. In fact, in the idealized case of both a completely incoherent illumination and an eikonal interaction the latter two functions are independent of $p$, so that the momentum integral is finite only if the phase vanishes identically, i.e., for
\begin{equation}
 \eta m + (\eta + 1) \frac{d}{d_1}n=0.
\end{equation}
It would imply that an interference pattern is observed only if $(\eta+1)d/(\eta d_1)$ is a rational number.

This strong resonance condition gets relaxed if we account for the weak momentum dependence of the remaining integrand in (\ref{eqn:interferencepattern1}). We assume that those index pairs $(m,n)$ of the double sum contribute appreciably where the phase variation in the momentum integration does not exceed about $\pi$. Since the value of $|p|/p_z$ is bounded by the angular spread $\alpha$ this leads to the relaxed condition 
\begin{equation}
 \left| \frac{\eta+1}{\eta} \frac{d}{d_1} - \frac{r}{s} \right| \leq 
 \frac{1}{2\alpha}\frac{d}{L},
 \label{eqn:TLresbound}
\end{equation}
Here $r$ and $s$ are natural numbers without common divisor. They indicate the type of resonance and specify the set of index pairs $(m,n)\in\{(r\ell,s\ell):\ell\in\mathbb{Z}\}$ contributing to the sum.

If $\alpha$ is about $1$\,mrad and $L/d=10^{5}$ the right hand side of (\ref{eqn:TLresbound}) is on the order of $10^{-2}$, and in practice only a single resonance dominates
for each set of parameters $d$, $d_1$, $\eta$. The expression for the interference pattern thus simplifies to
\begin{eqnarray}
 w_3 \left(x \right) &=& \frac{1}{G} \sum_{\ell=-\infty}^{\infty} A_{s\ell}^{*} \exp\left(2 \pi i\frac{ \ell x}{d_3}  \right)
\int \frac{\diff p}{p_z} \, D \left( \frac{p}{p_z} \right) \nonumber \\
&& \times \,  B_{r \ell } \left( s \ell \frac{d}{d_1}\frac{L}{L_{\textrm{T}}}; p, - s \ell \frac{d}{d_1} \right) .
\label{eqn:interferencepattern2}
\end{eqnarray} 
Here, the period of the interference pattern is given by 
\begin{equation}
 d_3 = \left( \frac{r}{d} - \frac{s}{d_1}\right)^{-1} .
\end{equation} 
Note that a large interference contrast is obtained for the low order resonances (with small values of the integers $r$ and $s$) because the Fourier and Talbot-Lau coefficients $A_m, B_m$ generically decrease with their order $m$.

The standard choice in experiments is to take equal gratings, $d=d_1$, in an equidistant configuration, $\eta=1$. This corresponds to the $r:s = 2:1$ resonance. One may obtain even larger contrasts with $\eta=1$ by using the basic $r:s=1:1$ resonance, at the expense of dealing with different gratings, $d_1=2d$.

Also magnifying or demagnifying interferometers can be realized \cite{Brezger2003a}. Consider, for example, the case where the first two gratings are equal, $d=d_1$. The period of the interference signal is then given by $d_3 = d / (r-s)$ and cannot thus be greater than $d$.
The condition $r/s= \left( \eta+1 \right) / \eta$ with $\eta >0$ implies that $r$ must be strictly greater than $s$. If one wishes to decrease $d_3$, one must choose $r-s \geq 2$ which requires $ \eta < 1$, i.e., a compressed setup. However, the values of $r$ and $s$ grow with decreasing $d_3$ so that the visibility of the resulting interference pattern decreases significantly. A magnifying interferometer, on the other hand, can only be realized with different gratings $d \neq d_1$. 

If the aim is to create a high contrast interference pattern with a specific period $d_3$ a low-order resonance should be taken, such as $r:s=2:1$ or $r:s=1:1$. The experimental setup parameters $\left(\eta,d,d_1 \right)$ must then satisfy the equations%
\begin{eqnarray}
	\left(\eta +1 \right) d &=& r d_3  \\
	\eta d_1 &=& s d_3
\end{eqnarray}
with desired resonance parameters $\left(r,s,d_3 \right)$. The solution is not unique, in general, so that one has a certain freedom to account for experimental limitations.

\subsection{The Talbot-Lau effect in eikonal approximation}
\label{sec:TLeikonal}

The expression for the interference pattern can be further simplified if the grating interaction is treated in eikonal approximation. The grating coefficients (\ref{eqn:gratingcoeff}) turn then into the momentum-independent Fourier coefficients $b_j$ of the transmission function (\ref{eq:bj}). Similarly, the generalized Talbot-Lau coefficients (\ref{eqn:TLcoeff}) then reduce to the basic coefficients given in Eq.~(\ref{eqn:Beik}) so that the prediction for the density pattern (\ref{eqn:interferencepattern1}) simplifies to the series 
\begin{eqnarray}
 w_3 \left(x \right) &=& \frac{1}{G} \sum_{\ell,m = -\infty}^{\infty}\, A_{\ell} \, B_{m} \left( \eta \left(m + \ell \frac{d}{d_1} \right) \frac{L}{L_{\textrm{T}}}\right) \nonumber \\
&& \times \, \widetilde{D} \left( \frac{2 \pi L}{d} \left( \eta m + (\eta + 1) \ell \frac{d}{d_1} \right) \right) \nonumber \\
&&\times \, \exp\left(2 \pi i \left( m + \ell \frac{d}{d_1} \right)\frac{ x}{d} \right) .
\label{eqn:w3_eik1}
\end{eqnarray}
It involves the Fourier transformation of the angular distribution, 
\begin{equation}
 \widetilde{D} \left( \omega \right) = \int \frac{\diff p}{p_z} D \left( \frac{p}{p_z}\right) e^{-i \omega p / p_z} .
 \label{eqn:Dtilde}
\end{equation} 
In the case of a very broad momentum distribution this characteristic function $\widetilde{D} $ can be replaced by a Kronecker-$\delta$ function, which yields the basic result \cite{Brezger2003a,Hornberger2004a}
\begin{eqnarray}
w_3 \left(x \right) &=& \frac{1}{G} \sum_{\ell=-\infty}^{\infty} A_{s\ell}^{*} B_{r \ell } \left( s \ell \frac{d}{d_1} \frac{L}{L_{\textrm{T}}} \right) \nonumber \\
&& \times \,  \exp\left(2 \pi i \ell\frac{ x}{d_3}  \right) .
\label{eqn:w3_eik2}
\end{eqnarray}
The comparison with Eq.~(\ref{eq:TE}) shows clearly that Talbot-Lau interference is based on the Talbot effect, both described by the Talbot-Lau coefficients defined in (\ref{eqn:Beik}).

It should be emphasized, though, that in order to establish that an observed fringe pattern is really due to a quantum effect one must compare the quantum prediction with the result of the corresponding classical calculation. This is necessary since a classical moir\'{e}-type shadow pattern might give rise to a similar observation, even tough the classical contrast is typically much smaller. It is shown in the appendix how the classical treatment can be formulated in the same framework as the quantum case using a phase-space description.

Finally, since Eq.~(\ref{eqn:w3_eik2}) assumes the resonance condition to be exactly met, it cannot be used to assess the adjustment precision required in the experiment.
As shown in the next section, one can evaluate the necessary precision in the eikonal approximation by explicitly taking into account the finite transverse momentum spread and the finite size of the signal detector.

\subsection{Adjustment requirements}

In order to allow for deviations from the exact resonance condition we use Eq.~(\ref{eqn:w3_eik1}) when evaluating the expression for the detection signal.
According to Eq.~(\ref{eqn:signal}) the detection signal $S(x_s)$ is then characterized by the Fourier coefficients
\begin{eqnarray}
 S_n &=& \frac{W}{G}A^{\prime *}_{n}  \sum_{\ell,m = -\infty}^{\infty}\, A_{\ell}  \, B_{m } \left( \eta \left(m + \ell \frac{d}{d_1} \right) \frac{L}{L_{\textrm{T}}}\right) \nonumber \\
&& \times \, \widetilde{D} \left( 2 \pi L \left( \frac{\eta}{d} m + \frac{\eta + 1}{d_1} \ell \right) \right) \nonumber \\
&&\times \, \sinc \left( \pi W \left( \frac{m}{d} + \frac{\ell}{d_1} - \frac{n}{d_3}\right)\right).
\label{eqn:signal2}
\end{eqnarray} 
Here, the $A'_n$ are the Fourier expansion coefficients of the third grating profile $\left|t_3 \left(x \right) \right|^2$ with the period $d_3$. Consider now small deviations $\left(\delta L, \delta \eta , \delta d_1, \delta d, \delta d_3 \right)$ from the setup parameters $\left(L, \eta , d_1, d, d_3 \right)$ satisfying a particular Talbot-Lau resonance condition $r:s$. The distances $L_1=L$ and $L_2=\eta L$ can thus vary independently. 
Instead of approximating the characteristic function $\widetilde{D}$ by a Kronecker-$\delta$, we account for the small deviations by the refined approximation $\widetilde{D} \left(\omega + \varepsilon \right) \approx \delta_{\omega,0} \widetilde{D} \left(\varepsilon \right)$. It holds as long as the main argument $\omega$ is either zero or much greater in modulus than the width $1/\alpha$, and as long as $\varepsilon < 1/\alpha$. 

One can thus split the argument of $\widetilde{D} $ in (\ref{eqn:signal2}) into the ideal resonance part $\omega$ and in the part $\varepsilon$ containing the parameter deviations.
The same procedure may be done with the $\sinc$ term, which is sharply peaked if the detector size $W$ exceeds the period $d$ by orders of magnitude.
If the small parameter deviations are taken into account to first order one obtains the same resonance relation 
for the summation indices in (\ref{eqn:signal2}) as in the ideal case, while the $n$th Fourier expansion coefficient of the signal (with respect to the period $d_3$) is multiplied by the reduction factor
\begin{eqnarray}
R_{n} &=& \widetilde{D} \left( \frac{2 \pi s n L}{d_1} \left[ \frac{\delta L}{L} - \frac{\delta \eta }{\eta } + \left(\eta +1 \right) \left( \frac{\delta d}{d} - \frac{\delta d_1}{d_1} \right) \right]\right) \nonumber \\
&&\times \, \sinc \left(\frac{\pi s n W}{d_1} \left[ \frac{\delta d_1}{d_1} - \frac{\delta d}{d} + \frac{1}{\eta} \left( \frac{\delta d_3}{d_3} - \frac{\delta d}{d} \right] \right) \right) . \nonumber\\
&&
\label{eqn:reductionfactor}
\end{eqnarray}
The absolute value of this factor is less than $1$ for $n \neq 0$ so that the contrast of the detection signal is effectively reduced by the deviations. In particular, the sinusoidal visibility (\ref{eqn:Vexp}) of the signal is reduced by the factor $R_1$.
Typical experiments \cite{Brezger2002a,Brezger2003a,Hackermuller2003a,Gerlich2007a} are characterized by $W/d \approx 10^2$, $L/d \approx 10^5$ and $\alpha \approx 10^{-3}$ so that already relative parameter deviations on the order of $0.1 \%$  strongly affect the interference contrast.

To obtain a simple and conservative estimate we treat all imprecisions as independent, specializing to the standard case $r:s=2:1$, where the grating distances and grating periods are equal. Bounds for the adjustment precisions are then obtained from Eq.~(\ref{eqn:reductionfactor}) by requiring the arguments to be smaller than the widths of the momentum distribution and of the $\sinc$ function, respectively,
\begin{eqnarray}
	\frac{\delta L}{L} + 2\frac{ \delta d}{d} &<& \frac{d}{4 \pi \alpha L} 
\end{eqnarray}
and
\begin{eqnarray}
	\frac{\delta d}{d} &<& \frac{d}{4W}.
\end{eqnarray}
This quantifies to what degree a better collimation of the beam and a smaller detector size relax the required adjustment precision, albeit at the expense of a loss of signal.

\section{Effect of the interaction potential} \label{sec:potential}

The preceding section showed how the general coherent state transformation effected by a diffraction grating enters the Talbot-Lau calculation via the generalized grating coefficients (\ref{eqn:gratingcoeff}). 
In general, this transformation is determined by the potential $V \left(x,z \right)$ due to the long-range dispersion forces acting on the beam particles while they pass the grating structure. 

Before we present a general way to account for the presence of this potential in Sect.~\ref{sec:semiclass} it is helpful to discuss the most important grating-particle interactions in a simpler form, by using the \emph{eikonal approximation}. It treats the grating as the combination of an absorption mask and a phase modification, and it can be characterized by a grating transmission function
\begin{equation}
 t \left(x \right) = \left|t\left(x \right) \right| \exp \left(- \frac{i m}{\hbar p_z} \int \diff z \, V \left(x,z \right) \right),
 \label{eqn:simpleeiko}
\end{equation}
whose amplitude $\left|t\left(\cdot \right) \right|\mapsto\{0,1\}$ describes the grating structure. 
Interaction-free gratings are modeled by a grating transmission function without phase, $t(x)=|t(x)|$, while pure phase gratings, such as the standing laser field, wave are characterized by $|t(x)|=1$.
The corresponding eikonal interference pattern is then obtained immediately
as described in Sect.~\ref{sec:TLeikonal}.

We proceed with a short overview of the typical grating potentials, discussing how they affect the interference contrast in the eikonal approximation. By convention, the diffraction grating is located at the longitudinal position $z=0$, as indicated in Fig.~\ref{fig:TLsetup}.

\subsection{Material gratings} \label{sec:potential_mat}

A neutral particle located within the slit of a material grating experiences a potential determined mainly by the attractive dispersion forces due to the grating walls. Other forces, such as the exchange interaction or the electro-static attraction due to a permanent dipole, are much less important for interferometry because they either occur only at very close distances or because they are diminished by rotational averaging. 
In any case, for all positions within a slit the walls are well approximated by an infinite surface in the $yz$-plane \cite{Grisenti1999a,Bruhl2002a}. The grating potential is thus set to be solely $x$-dependent and acting only within the time of passage $t=m b /p_z$ throughout the grating of thickness $b$, while it is set to be zero outside.

In general, the dispersion force between a polarizable particle and a material plane is described by the expression of Casimir and Polder \cite{Casimir1948a} and its generalizations, e.g.~\cite{Wylieboth,Buhmann2005a}. In the close distance limit it reduces to the van der Waals potential $V (x) = -C_3/x^{3}$, with $x$ the distance to the surface, and at large distances, where retardation plays a role, one has the asymptotic form $V(x) = -C_4/x^{4}$. 
The interaction constants $C_3,C_4 > 0$ are determined by the frequency dependent polarizability of the particle (as well as the dielectric function of the grating material), and the regime of validity of the limiting forms is delimited by the wavelengths corresponding to the strong electronic transitions \cite{Derevianko1999a,Bruhl2002a,Madronero2007a}.

In any case, the dispersion force diverges on the grating walls, rendering 
the eikonal approximation invalid in close vicinity to the walls. 
Since the contributions of these regions do not alter the interference contrast appreciably, but lead to 
numerical noise, 
we will discuss a reasonable criterion to blind out the beam close to the grating walls in Section \ref{sec:semiclass}.

\begin{figure}
\includegraphics[width=8cm]{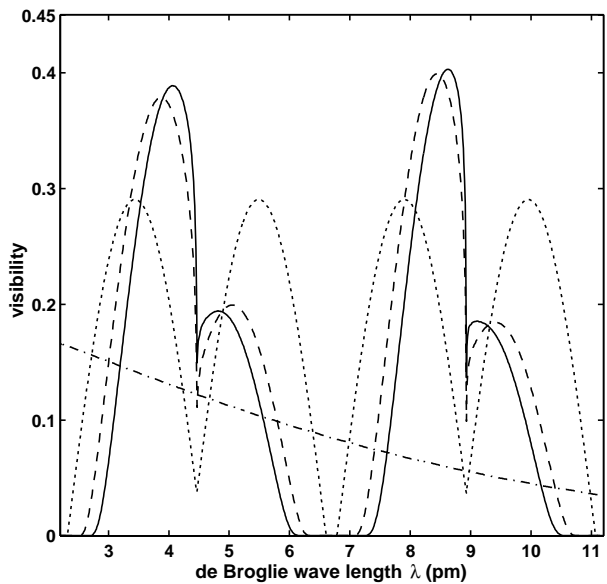}
\caption{\label{fig:lambda_mat} Talbot-Lau visibilities 
(\ref{eqn:Vexp}) for the equidistant setup with three material gratings \cite{Brezger2002a} as a function of the de Broglie wavelength 
of the interfering C$_{70}$ fullerenes.  At the second grating we assume no dispersive interaction (dotted line), a van der Waals interaction (dashed line), and a retarded van der Waals interaction (solid line). The dash-dotted line represents the classical calculation with a retarded interaction. It depends on the fictitious ``wavelength'' $\lambda=h/(m v_z)$ via the velocity $v_z$ because the interaction time is velocity dependent.
}
\end{figure}

Figure \ref{fig:lambda_mat} shows the interference visibility (\ref{eqn:Vexp}) for a typical Talbot-Lau experiment with C$_{70}$ fullerene molecules of mass $m=840 \, \text{amu}$ \cite{Brezger2002a},
plotted versus their de Broglie wavelength. All gratings are assumed to be separated by the distance $L=22 \, \text{cm}$ and to be made of gold with a period of $d=991 \, \text{nm}$, a slit width of $476 \, \text{nm}$, and a thickness of $b=500 \, \text{nm}$. One can see two peaks corresponding to the first and second Talbot order, at $\lambda=4.5\,\text{pm}$ and at $\lambda=9.0\,\text{pm}$, respectively. The solid line represents the eikonal approximation with a retarded asymptotic potential ($C_4=3\hbar c \alpha_0 / 8 \pi$ with the static polarizability $\alpha_0 = 96.7\,$\AA$^3$ obtained via the Clausius-Mossotti relation \cite{Compagnon2001a}),
while the dashed and the dotted lines correspond to the van der Waals potential ($C_3=10\, \text{meV}\,\text{nm}^3$) and to the absence of an intra-slit potential, respectively. One observes that the presence of the dispersion forces changes the interference characteristics significantly. The asymmetry in the double-peak structure compared to the interaction-free case is due to the fact that the particles with a larger velocity, i.e., with a smaller wavelength $\lambda$, receive a smaller eikonal phase than the slower particles.
Moreover, note that the effect of retardation has a small but visible influence on the fringe contrast.

The dash-dotted line in Fig.~\ref{fig:lambda_mat} represents the moir\'{e}-like effect as expected from classical mechanics. It was calculated in a phase space formulation as described in the appendix,
using the classical correspondence of the eikonal approximation with the $C_4$-potential. As one expects, the dependence of the classical result on the fictitious ``wavelength'' $\lambda=h/(m v_z)$, which is due to the velocity dependence of the classical deflection, differs strongly from the quantum results. 

We emphasize that all numerical results presented in this paper are obtained for a fixed longitudinal velocity  $v_z = h / (\lambda m)$ of the particles. A comparison with the experiment still requires the results to be averaged with respect to the velocity distribution in the beam. This may pose a severe restriction when using particles with larger polarizability because the increased dispersive interaction decreases the width of the double-peaks seen in Fig.~\ref{fig:lambda_mat}  significantly \cite{Gerlich2007a}. One way to avoid this is to replace the material diffraction grating by a standing laser wave.

\subsection{Laser gratings}

Recently, a Kapitza-Dirac-Talbot-Lau interferometer (KDTLI) was demonstrated, where the central grating is formed by a standing light wave \cite{Gerlich2007a}.
In the eikonal approximation the phase of an incoming plane wave is modulated according to the potential created by the off-resonant interaction with the standing laser beam. It is determined by the energy of the induced electric dipole in the oscillating field, and is therefore proportional to the laser power $P_L$ and to the dynamic polarizability  $\alpha_{\omega}$ of the beam particles at the laser frequency,
\begin{equation}
 V \left( x,z \right) = - \frac{4 P_L \alpha_{\omega}}{\pi \varepsilon_0 c w_y w_z} \sin^2 \left( \pi \frac{x}{d}\right) e^{-2 z^2 / w_z^2} .
 \label{eqn:laserpot}
\end{equation} 
Here, $w_y$ and $w_z$ are the waists of the Gaussian mode, and $w_y$ is chosen large compared to the detector size in order to guarantee a regular grating structure. We therefore disregard the $y$-dependence by setting $y=0$. Moreover, for sufficiently small $w_z$ an effective one-dimensional treatment of the potential is permissible, where one replaces the $z$-dependence by a parametric time dependence $z=p_z t / m$, as discussed below in Sect.~\ref{sec:pertexp}.

The $z$-integration in (\ref{eqn:simpleeiko}) renders the eikonal phase independent of $w_z$, which already indicates that the elementary eikonal approximation  will cease to be valid if the laser waist is increased. Nonetheless, it yields an interference contrast that fits well to the measured data of the recent fullerene experiments \cite{Gerlich2007a}. Moreover, it admits a simple, closed expression for the Talbot-Lau coefficients (\ref{eqn:Beik}),
\begin{equation}
	B_m \left(\xi \right) = J_m \left( - \frac{4 M P_L \alpha_{\omega} }{\sqrt{2\pi} \hbar c \varepsilon_0 w_y p_z} \sin \pi \xi \right),
	\label{eqn:Beiklaser}
\end{equation}
where $M$ is the mass of the beam particles and $J_m$ stands for the $m$th order Bessel function of the first kind \cite{Abramowitz1965a}.

\begin{figure}
\includegraphics[width=8cm]{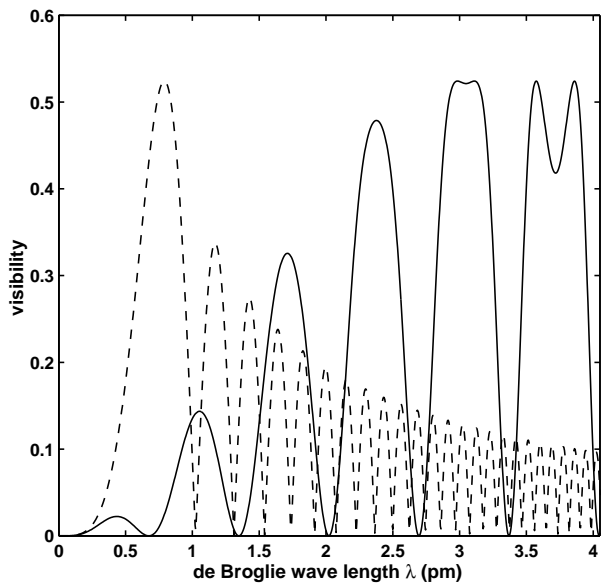}
\caption{\label{fig:lambda_laser} Fringe visibilities (\ref{eqn:Vexp}) for a Kapitza-Dirac-Talbot-Lau interferometer \cite{Gerlich2007a} in the eikonal approximation. The solid line gives the quantum result for C$_{70}$ fullerenes, while the dashed line represents the corresponding classical calculation (neglecting photon absorption). Here we assume a laser power of $P_\text{L}=6\,$W, a vertical waist of  $w_y=900\,\mu\text{m}$, a grating distance of $L=105 \, \text{mm}$, a grating period of  $d=266 \, \text{nm}$, and a dynamic polarizability $\alpha_{\omega} = 118 \,$\AA$^3$ corresponding to a laser wavelength of $532 \, \text{nm}$.}
\end{figure}

The solid line in Figure \ref{fig:lambda_laser} shows the visibility (\ref{eqn:Vexp}) for the Viennese KDTLI with fullerenes  \cite{Gerlich2007a} as obtained from (\ref{eqn:Beiklaser}). In contrast to Figure \ref{fig:lambda_mat}, the visibility drops to zero whenever the wavelength corresponds to an integer multiple of the Talbot condition. This is explained by the fact that in 
the elementary Talbot effect (\ref{eqn:Talboteffect}) a pure phase grating with no absorptive walls leads to a constant density at multiples of the Talbot length. 
At the same time, there are broad regions of high contrast. They render the KDTLI setup superior to a material grating interferometer for particles with a high polarizability and a substantial velocity spread. Moreover, the interaction strength and the grating period can be tuned in a KDTLI rendering the laser grating the preferable choice to explore the validity regime of the eikonal approximation, as done in Section \ref{sec:numerics}.

The dashed line in Figure \ref{fig:lambda_laser} represents the classical version of the eikonal calculation as described in the appendix. 
Note that it differs significantly from the quantum result in the experimentally relevant wavelength regime, while the curves become indistinguishable for $\lambda \rightarrow 0$.
Finally, it should be emphasized that a realistic description of the laser grating must also account for the possibility of photon absorption. The effect of the resulting transverse momentum kicks
can be incorporated into the eikonal approximation by replacing the grating transformation kernel by a probabilistic sum of such kernels \cite{Gerlich2007a}.

\section{Semiclassical approach to the grating interaction} \label{sec:semiclass}

The de Broglie wavelength is by far the smallest length scale in the matter wave interference experiments considered in this paper.
It seems therefore natural to use a semiclassical approach for calculating the grating transformation used in (\ref{eq:Undef}). We will show how the elementary eikonal approximation (\ref{eqn:simpleeiko}) can be derived from the appropriate semiclassical formulation if a high-energy limit is taken. Moreover, one can obtain a refined eikonal approximation, where an incoming plane wave is still merely multiplied by a factor. Since this factor depends on the transverse momentum of the incoming wave it renders the numerical implementation more elaborate.

However, the main result of this section is the expression (\ref{eqn:semiclassresult})--(\ref{eqn:semiclassphase}) for the scattering factor which approximates the scattering transformation of a transverse plane wave in the semiclassical high-energy regime. As such it permits to evaluate the generalized grating coefficients (\ref{eqn:gratingcoeff}) straightforwardly, see Eq.~(\ref{eqn:gratingcoeff2}). As shown in Section \ref{sec:numerics}, this result outperforms the eikonal approximation and it may serve to characterize its regime of validity.
The derivation is based on the semiclassical approximation of the two-dimensional time evolution operator $\oU_t$ determined by the grating interaction potential $V \left(\vx \right)=V \left(x,z \right)$.

\subsection{The scattering factor}

We proceed to incorporate the grating interaction
by means of the formalism of scattering theory \cite{Taylor1972a}. Its basic tool is the scattering operator defined as
\begin{equation}
 \oS = \lim_{t \rightarrow \infty} \oU_{-t}^{(0)} \oU_{2t} \oU_{-t}^{(0)},
 \label{eqn:Soperator}
\end{equation} 
with $\oU_{t}^{(0)}=\exp \left(-i \oP^2 t /2 m \hbar \right)$ the free time evolution operator
for the motion in the $xz$-plane
 and $\oU_{t}$ the complete time evolution operator, which includes the grating potential $V(\vx)$. It is pertinent to use $\oS$ instead of $\oU_t$ since the scattering operator transforms the state instantaneously leaving the asymptotic dynamics to be described by the free time evolution. Since the latter is easily incorporated using Wigner functions this fits to the phase space description of Sect.~\ref{sec:generalTL}, where the initial beam state entering a Talbot-Lau interferometer is propagated freely to the second grating before the scattering transformation is applied. In the expression  (\ref{eqn:generalprop}) for the grating propagator, which serves to calculate the interference pattern (\ref{eqn:interferencepattern2}), one may thus use the S-matrix (\ref{eqn:Soperator}) in place of the general unitary operator $\oU$. 

In the basis of the improper plane wave states $\la \vx|\vp \ra=(2\pi\hbar)^{-1}\exp(i\vx\cdot\vp/\hbar)$ the scattering operator (\ref{eqn:Soperator}) is conveniently described by the scattering factor
\begin{equation}
 \Phi \left( \vx, \vp \right) = \frac{\la \vx | \oS | \vp \ra}{\la \vx | \vp \ra}
 \label{eqn:Sfactor}
\end{equation} 
Notice that $\oS$ acts in the Hilbert space defined on the two-dimensional plane $\vx = \left(x,z \right)$. However, as will be justified below, the longitudinal motion may be separated and treated classically
so that the transformation is confined to the transverse dimension, while the $z$-coordinate turns into an effective time coordinate for a given longitudinal momentum $p_z$. The brings about the 
reduced scattering factor
\begin{equation}
\phi \left(x,p \right)=\Phi \left( x\ex, p\ex + p_z \ez \right),
\label{eqn:Sfactor_1D}	
\end{equation}
which depends parametrically on $p_z$ and which is evaluated at the position $z=0$ of the center of the diffraction grating.

The reduced scattering factor $\phi \left(x,p \right)$ describes the phase and amplitude modification of a transverse plane wave with momentum $p$ due to the grating interaction. It enters the Talbot-Lau calculation via the Fourier coefficients (\ref{eq:Undef}) after an expansion in the plane wave basis. It follows that the generalized grating coefficients (\ref{eqn:gratingcoeff}) are directly related to $\phi\left(x,p \right) $ by a Fourier transformation,
\begin{equation}
 b_n \left( p\right) = \frac{1}{d} \int_{-d/2}^{d/2} \diff x \, e^{-2 \pi i n x /d} \phi \left( x, p \right).
 \label{eqn:gratingcoeff2}
\end{equation}
The calculation of the Talbot-Lau interference thus reduces to evaluating the scattering factor (\ref{eqn:Sfactor_1D}).

\subsection{Semiclassical calculation}

We proceed to calculate the scattering factor (\ref{eqn:Sfactor_1D}) by means of the semiclassical asymptotic approximation. It assumes the action of those trajectories through the interaction region, which contribute to the path integral for its position representation, to be much larger than $\hbar$, and the particle wavelength to be much smaller than the scale where the interaction potential changes appreciably,  $\left| \lambda \nabla V \right| \ll V$). 
These conditions are well satisfied in the typical experimental situation if we disregard the regions very close to the gratings where the potential exceeds the kinetic energy.
We may thus approximate the time evolution operator in the position representation by the semiclassical van Vleck-Gutzwiller propagator \cite{Gutzwiller1967a}
\begin{eqnarray}
 \la \vx | \oU_t |\vx_0 \ra &=& \frac{1}{2 \pi i \hbar} \sqrt{\left| \det \left( \frac{\partial^2 S_{t} \left( \vx, \vx_0\right)}{\partial \vx \partial \vx_0}\right)\right|} \nonumber \\
&&\times \, \exp \left(\frac{i}{\hbar} S_{t} \left( \vx, \vx_0\right)\right) .
\label{eqn:VVG1}
\end{eqnarray}
Here, $S_{t}$ is the  action  of the classical trajectory travelling during time $t$ from the position $\vx_0$ to $\vx$. In general, there might be more than one such trajectory, which would require taking special care of almost coalescing trajectories and of the associated Morse index. However, in the interferometric setup we are in the high-energy regime, where the interaction potential is much weaker than the energy of the incoming particles, $|V| \ll E$.
All relevant contributions are therefore characterized by a single, slightly deflected classical trajectory passing the interaction region. 

It is now convenient to specify this trajectory in terms of the deviation from the undeflected straight line, as specified by the initial position $\vx_0$ and momentum $\vp_0$. The momentum change after time $t$ is given by
\begin{eqnarray}
\Delta \vp_t \left(\vx_0, \vp_0 \right) &=& -\int_0^t \diff \tau \, \nabla V \left( \vx_{\tau} \left(\vx_0, \vp_0 \right) \right),
\label{eq:Deltap}
\end{eqnarray} 
so that the deflected trajectory reads
\begin{eqnarray}
\vx_{t} \left(\vx_0, \vp_0 \right) &=& \vx_0 + \frac{t}{m}\vp_0+\int_0^t \frac{\diff \tau}{m} \, \Delta\vp_{\tau} \left(\vx_0, \vp_0 \right) \label{eqn:trajx} \\
\vp_{t} \left(\vx_0, \vp_0 \right) &=& \vp_0 + \Delta \vp_t \left(\vx_0, \vp_0 \right). \label{eqn:trajp}
\end{eqnarray} 
In the van Vleck-Gutzwiller propagator (\ref{eqn:VVG1}) the contributing trajectory is specified by the boundary values $\vx_0$ and $\vx$. For the following calculation it is important to rewrite it as an initial value problem, specified by the initial phase space point $\left(\vx_0, \vp_0 \right)$ of the trajectory \cite{Heller1991a}.
\begin{eqnarray}
 \la \vx | \oU_t |\vx_0 \ra &=& \frac{1}{2\pi i \hbar} \int \diff^2 p_0 \, \sqrt{\left| \det \left( \frac{\partial \vx_{t} \left(\vx_0, \vp_0 \right)}{\partial \vp_0} \right) \right|}
\nonumber\\
&&\times \delta \left( \vx - \vx_{t} \left(\vx_0, \vp_0 \right) \right) \exp \left( \frac{i}{\hbar} S_{t} \left( \vx_0, \vp_0\right) \right)
\nonumber\\&& \label{eqn:VVG2}
\end{eqnarray}
Plugging this into the expression (\ref{eqn:Soperator}) for the 2d scattering operator yields a semiclassical approximation for the scattering factor (\ref{eqn:Sfactor}).
\begin{align}
	\Phi \left( \vx, \vp \right) =& \lim_{T\rightarrow \infty} \frac{m}{T} \iint \frac{\diff^2 r_0 \diff^2 p_0}{\left(2 \pi \hbar \right)^2} \, \sqrt{\left| \det \left( \frac{\partial \vx_{2T} \left(\vx_0, \vp_0 \right)}{\partial \vp_0} \right) \right|} 
\nonumber\\
&\times\exp \left( \frac{i}{\hbar} \Theta_{2T} \left( \vx_0, \vp_0 \right) \right)
	\label{eqn:Sfactor2}
\end{align}
The phase of the integrand is given by the action-valued function
\begin{eqnarray}
	\Theta_{2T} \left( \vx_0, \vp_0 \right) &=& S_{2T} \left( \vx_0, \vp_0\right) - \frac{m}{2T} \left( \vx - \vx_{2T} \left( \vx_0, \vp_0 \right) \right)^2 
\nonumber\\
&&- \vp \cdot \left( \vx - \vx_0 - \frac{\vp}{2m} T \right),
	\label{eqn:Sphase1}
\end{eqnarray}
while the amplitude is determined by the stability determinant of the associated trajectory.

The semiclassical van Vleck-Gutzwiller propagator (\ref{eqn:VVG1}) used here is a stationary phase approximation of the  time evolution operator in path integral formulation \cite{Gutzwiller1967a}. To remain at a consistent level of approximation it is therefore necessary to evaluate the phase space integral in (\ref{eqn:Sfactor2}) in the stationary phase approximation as well \cite{Bleistein1975a}.
The stationary point of the phase $\Theta_{2T}$ is determined by the condition 
\begin{equation}
\left( \left. \frac{\partial \Theta_{2T}}{\partial \vx_0} \right|_{\vp_0}, \left. \frac{\partial \Theta_{2T}}{\partial \vp_0} \right|_{\vx_0} \right) = 0	.
\end{equation}
Noting the initial value derivatives of the classical action 
\begin{eqnarray}
	\left(\left. \frac{\partial S_{t}}{\partial \vx_0} \right|_{\vp_0}\right)^T &=& - \vp_0 +\left( \left. \frac{\partial \vx_t}{\partial \vx_0} \right|_{\vp_0}\right)^{\rm T} \vp_{t} \\
	\left(\left. \frac{\partial S_{t}}{\partial \vp_0} \right|_{\vx_0}\right)^T &=& \left(\left. \frac{\partial \vx_t}{\partial \vp_0} \right|_{\vx_0} \right)^{\rm T}\vp_{t},
\end{eqnarray}
and using the fact that the matrix $\partial\vx_{2T}/\partial\vp_0$ is invertible for our trajectories,
the stationary phase condition leads to the equations
\begin{eqnarray}
	\vx &=& \vx_{2T} \left( \vx_0, \vp_0 \right) - \frac{T}{m} \vp_{2T} \left( \vx_0, \vp_0 \right) \label{eqn:statx}\\
	\vp &=& \vp_0 \label{eqn:statp}.
\end{eqnarray}
They serve to determine the initial position $\vx_0(\vx,\vp)$ of the trajectory implicitly. Using the general formula for a four-dimensional integral, e.g. \cite[Eq. (A.30)]{Hornberger2002b},  the 2d scattering factor (\ref{eqn:Sfactor2}) then takes the form
\begin{equation}
	\Phi \left( \vx, \vp \right) = \lim_{T\rightarrow \infty} A_{2T} \left( \vx_0 \left( \vx, \vp \right), \vp \right) \exp \left( \frac{i}{\hbar} \Theta_{2T} \left( \vx_0 \left( \vx, \vp \right),\vp \right) \right).
	\label{eqn:Sfactor3}
\end{equation}
The amplitude modification 
\begin{equation}
A_{2T} \left( \vx_0, \vp_0 \right)	= \left| \det \left( \frac{\partial \vx_{2T} \left( \vx_0, \vp_0 \right)}{\partial \vx_0} - \frac{m}{T} \frac{\partial \vp_{2T} \left( \vx_0, \vp_0 \right)}{\partial \vx_0} \right) \right|^{-\frac{1}{2}}
\label{eqn:Samp1}
\end{equation}
is obtained, in a tedious but straightforward calculation, by using the Poisson relation between the conjugate variables $\vx_t$ and $\vp_t$ of the trajectory \cite{LandauM} when evaluating the product of two determinants. The matrix of derivatives in (\ref{eqn:Samp1}) can be computed by taking the initial value derivative of the equation of motion for the trajectory and solving the resulting ordinary differential equation. 

\begin{figure}
\includegraphics[width=7.5cm]{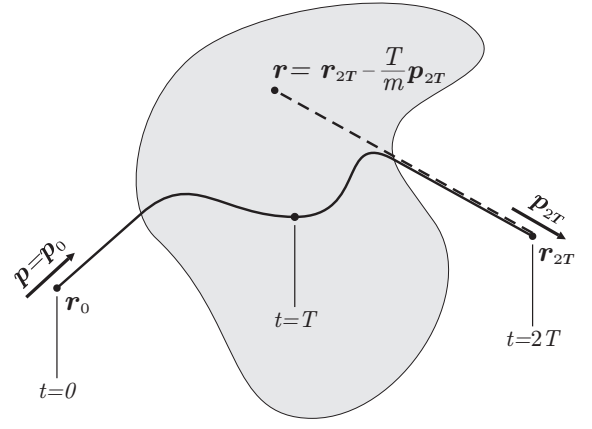}
\caption{\label{fig:scattskizze} Sketch of the stationary phase conditions for the semiclassical scattering factor $\Phi \left( \vx, \vp \right) $ defined in (\ref{eqn:Sfactor3}). While the momentum $\vp$ fixes the asymptotic initial momentum $\vp_0$ of the classical trajectory, the initial position $\vx_0$ is determined implicitly by the spatial point $\vx$. The latter defines the asymptotic final position $\vx_{2T}$ of the trajectory by means of a free evolution determined by the asymptotic final momentum $\vp_{2T}$. For interaction potentials that admit a scattering theory description this construction guarantees that the action (\ref{eqn:Sphase1}) and the stability amplitude (\ref{eqn:Samp1}) converge as $T\to\infty$. The range of the scattering potential is indicated by the shaded area.}
\end{figure}

Figure \ref{fig:scattskizze} shows how the stationary point can be understood from the point of view of a classical scattering trajectory. Given the phase space coordinates $\vx, \vp$ 
determining the matrix element of the scattered plane wave, 
the momentum $\vp$ fixes the initial momentum of the classical trajectory passing through the interaction region. The position $\vx$ determines the final position $\vx_{2T}$ of the deflected trajectory, which is obtained after a free motion during time $T$ in a direction given by the final momentum $\vp_{2T}$ of the trajectory. This free evolution ensures that the associated action (\ref{eqn:Sphase1}) and stability amplitude (\ref{eqn:Samp1}) is independent of $T$ in the limit $T \rightarrow \infty$. 
Note that Fig.~\ref{fig:scattskizze} overemphasizes the effect of the potential, since in all interferometrically relevant situations the deflection of the classical trajectories will be only a
small correction to the free rectilinear path.

The limit in the semiclassical result (\ref{eqn:Sfactor3}) is obtained already at a finite time $T$ if the interaction potential may be considered to have a finite range. The relevant scale is the grating passage time, given by $t=m b /p_z$ for material gratings and $t = m w_z / p_z$ for laser gratings, respectively. This follows from the fact that outside of the grating potential the classical trajectories remain rectilinear, so that neither the stationary phase condition (\ref{eqn:statx}) nor the stationary phase $\Theta_{2T}$ and amplitude $A_{2T}$ are affected by a further increase in $T$.

\subsection{Iterative solution in the momentum deflection}
\label{sec:pertexp}

We will now give explicit approximate solutions for the semiclassical scattering factor (\ref{eqn:Sfactor3}) which are valid in the high-energy limit, i.e., whenever the longitudinal kinetic energy $p_z^2/2 m$ of the incoming particles is large compared to the interaction potential $|V(\vx)|$. 
In general, one has to solve Eq.~(\ref{eqn:statx}) for the initial value $\vx_0$. Rewriting it in terms of the expressions (\ref{eqn:trajx}) and (\ref{eqn:trajp}) for the positions and momenta of the trajectory, one obtains an implicit equation for $\vx_0(\vx,\vp)$ which may be solved iteratively.
\begin{align}
	\vx_0 \left( \vx, \vp \right) =& \vx - \frac{T}{m} \vp + \frac{T}{m} \Delta \vp_{2T} \left( \vx_0\left( \vx, \vp \right), \vp \right)
\nonumber\\
& -  \int_{0}^{2T} \frac{\diff t}{m} \, \Delta \vp_t \left( \vx_0\left( \vx, \vp \right), \vp \right)
	\label{eqn:trajexpansion}
\end{align}

\subsubsection{The Glauber eikonal approximation}

In the semiclassical short interaction time and high-energy limit, which applies to typical interference experiments, the transverse momentum deflection $\Delta \vp_{2T}$ is so weak that its contribution
can be neglected. 
This corresponds to the zeroth order solution of Eq.~(\ref{eqn:trajexpansion}),
\begin{equation}
	\vx_0^{(0)} \left( \vx, \vp \right) = \vx - \frac{T}{m} \vp.
\label{eqn:reihe0}
\end{equation}
In order to keep the approximation consistent 
one has to neglect the deflection in the expressions for the phase modification (\ref{eqn:Sphase1}) and for the amplitude modification (\ref{eqn:Samp1}) as well, by setting the classical trajectory to be the free rectilinear path. Consequently, there is no amplitude modification and the scattering factor (\ref{eqn:Sfactor3}) reads
\begin{equation}
	\Phi \left( \vx, \vp \right) = \exp \left( - \frac{i}{\hbar} \int_{-\infty}^{\infty} \diff t \, V \left( \vx + \frac{\vp}{m}t \right) \right).
\end{equation}
This is the \textit{Glauber eikonal approximation} obtained by R.~J.~Glauber from the Lippmann-Schwinger equation in a quite different argumentation \cite{Glauber1959a}. The reduced scattering factor (\ref{eqn:Sfactor_1D}) for the transverse dimension then reads
\begin{equation}
 \phi \left(x,p \right) = \exp \left( - \frac{im}{\hbar p_z} \int_{-\infty}^{\infty} \diff z \, V \left(x + \frac{p}{p_z}z , z \right)\right).
 \label{eqn:glauber}
\end{equation} 
The elementary eikonal approximation (\ref{eqn:simpleeiko}) follows from this expression in the limit of vanishing transverse momentum, $p \rightarrow 0$. It applies in the case of a well-collimated beam and a small grating thickness $b$,
i.e., $\left| p b / p_z \right| \ll d$.

\subsubsection{The deflection approximation} 

The Glauber eikonal approximation ceases to be valid as the interaction strength or interaction time increases, and one has to go to the first order in the momentum deflection $\Delta \vp_{t}$ when evaluating the stationary initial value (\ref{eqn:trajexpansion}),
\begin{eqnarray}
	\vx_0^{(1)} \left( \vx, \vp \right)& = &\vx - \frac{T}{m} \vp + \frac{T}{m} \Delta \vp_{2T} \left( \vx - \frac{T}{m} \vp, \vp \right)
\nonumber\\
&& -  \int_{0}^{2T} \frac{\diff t}{m} \, \Delta \vp_t \left( \vx - \frac{T}{m} \vp, \vp \right).
\label{eqn:reihe1}
\end{eqnarray}
The higher order terms, which are neglected here, involve derivatives of the momentum deflection $ \Delta \vp_{2T}$.

The time-evolved trajectory starting from this initial value is approximated, again to first order in $\Delta \vp_{t}$, by
\begin{eqnarray}
\label{eqn:1stordertraj}
\vx_t \left(  \vx_0^{(1)} (\vx, \vp), \vp \right)& \approx& \vx_t \left( \vx - \frac{T}{m} \vp, \vp \right)
\nonumber\\
&& + \frac{T}{m} \Delta \vp_{2T} \left( \vx - \frac{T}{m} \vp, \vp \right)
\\
&& - \int_{0}^{2T}  \frac{\diff \tau }{m}\, \Delta \vp_\tau \left( \vx - \frac{T}{m} \vp, \vp \right).
\nonumber
\end{eqnarray}
This trajectory must be used when calculating the action (\ref{eqn:Sphase1}) in order to ensure that all expressions are evaluated to the same order in the deflection $\Delta \vp_{t}$. Since this holds also for the time integral $\int \diff t \, V \left( \vx_t \left( \vx_0, \vp \right) \right)$  a $1$st order Taylor expansion of the potential is required. In total this yields
\begin{widetext}
\begin{eqnarray}
	\Theta_{2T} \left( \vx, \vp \right) &\approx& -\int_0^{2T} \diff t \,V \left( \vx_t \left( \vx - \frac{T}{m} \vp, \vp \right) \right) + \frac{1}{2m} \int_0^{2T} \diff t \, \Delta \vp_t^2 \left( \vx - \frac{T}{m} \vp, \vp \right) \nonumber \\
	&&-  \Delta \vp_{2T} \left( \vx - \frac{T}{m} \vp, \vp \right) \cdot \int_0^{2T} \frac{\diff t}{m} \, \Delta \vp_t \left( \vx - \frac{T}{m} \vp, \vp \right) + \frac{T}{2m} \Delta \vp_{2T}^2 \left( \vx - \frac{T}{m} \vp, \vp \right).
	\label{eqn:Sphase2}
\end{eqnarray}
On the other hand, when evaluating the amplitude (\ref{eqn:Samp1}) the zeroth order solution of the initial value  (\ref{eqn:reihe0}) must be used instead of (\ref{eqn:reihe1})   because the classical equation of motion for the matrix of initial value derivatives $(\partial \vx_{t} / \partial \vx_0, \partial \vp_{t} / \partial \vx_0 )$ is governed by the derivative of the interaction force rather than by the force itself.
Accounting for the modification (\ref{eqn:reihe1}) in calculating the stability determinant would therefore amount to a higher order correction in the momentum deflection. It follows that the amplitude factor is consistently approximated by
\begin{equation}
	A_{2T} \left( \vx, \vp \right)	\approx \left| \det \left( \frac{\partial \vx_{2T} \left( \vx - \vp T / m, \vp \right)}{\partial \vx_0} - \frac{T}{m} \frac{\partial \vp_{2T} \left( \vx - \vp T / m, \vp \right)}{\partial \vx_0} \right) \right|^{-1/2}.
	\label{eqn:Samp2}
\end{equation}
\end{widetext}
\subsubsection{Separation of the longitudinal motion} \label{sec:seplongmotion}

Based on the 
deflection approximation of the stationary scattering factor we may now derive a reduced scattering factor (\ref{eqn:Sfactor_1D}) that can be incorporated into the one-dimensional Talbot-Lau calculation and that significantly extends the validity regime of the eikonal approximation.
It is necessary for this purpose that the longitudinal motion remains unaltered. To a very good approximation this is indeed the case for the small trajectory deflections required above in the 
deflection approximation. This is due to the energy conservation during the scattering process, ensuring $(\vp + \Delta \vp_{2T})^2=\vp^2$ where $\vp= \left(p,p_z \right)$ is the incoming momentum and $\Delta \vp_{2T}$ the total deflection (\ref{eq:Deltap}). It follows that a small transverse deflection $\Delta p_{2T} \ll p_z$ yields in a well-collimated beam, $|p| \ll p_z$, a total longitudinal deflection $\Delta p_{z,2T} \approx p/p_z \Delta p_{2T}$ which is much smaller than $ \Delta p_{2T}$. 
The longitudinal part of the classical trajectory (\ref{eqn:1stordertraj}) may thus be treated as a free motion with constant momentum $p_z$,
\begin{equation}
	z_t \left( \vx_0^{(1)} \left( \vx, \vp \right), \vp \right) = z + \frac{p_z}{m} \left(t-T \right) + 
\Order \left( \Delta p_{2T}\frac{p}{p_z} \right).
\label{eqn:zfree}
\end{equation}
This reduces all the vectorial quantities in the 
semiclassical phase and amplitude modification to the transverse scalar quantities. 

In addition, one can now remove the longitudinal motion altogether from the scattering description by switching into a comoving frame with velocity $v_z=p_z/m$.  It is convenient to redefine the transverse components of the classical transverse trajectory so that they start at $-T$,
\begin{eqnarray}
	\Delta \bar{p}_t \left( x,p \right) &=& - \int_{-T}^t \diff \tau \, \partial_x V \left( \bar{x}_{\tau} \left( x,p \right), \frac{p_z}{m}\tau \right) \\
	\bar{x}_t \left( x, p \right) &=& x + \frac{p}{m}t -  \int_{-T}^t \frac{\diff \tau}{m} \, \Delta \bar{p}_{\tau} \left(x,p \right) \\
	\bar{p}_t \left(x,p \right) &=& p + \Delta \bar{p}_t \left(x,p \right).
\end{eqnarray}
These are the scalar analogues of (\ref{eq:Deltap})--(\ref{eqn:trajp}) but for the shift in the time coordinate by $-T$, which accounts for the asymptotic initial condition of the free longitudinal trajectory (\ref{eqn:zfree}).  

We can now state the result for the reduced scattering factor $\phi \left(x,p \right)=\Phi \left( x\ex, p\ex + p_z \ez \right)$ to first order in the momentum deflection.
\begin{equation}
\phi \left(x,p \right) = a \left(x,p \right) \exp \left(\frac{i}{\hbar} \theta \left(x,p \right) \right)
\label{eqn:semiclassresult}
\end{equation}
The amplitude is given by
\begin{equation}
	a \left( x, p \right) = \lim_{T \rightarrow \infty} \left|  \partial_x\bar{x}_T  \left(x,p \right) - \frac{T}{m} \partial_x\bar{p}_T\left(x,p \right)   \right|^{-\frac{1}{2}},
	\label{eqn:semiclassamp}
\end{equation}
%
while the phase can be simplified as
\begin{widetext}
\begin{eqnarray}
\theta \left( x, p \right) = -\lim_{T\to\infty} \int_{-T}^{T} \diff t \left[ V \left( \bar{x}_t(x,p) , \frac{p_z}{m}t \right) - \frac{t}{m}  \partial_x V \left( \bar{x}_t(x,p) , \frac{p_z}{m}t \right) \int_{t}^{T} \diff \tau \, \partial_x V \left( \bar{x}_\tau(x,p) , \frac{p_z}{m}\tau \right) \right] .
\label{eqn:semiclassphase}
\end{eqnarray}
\end{widetext}
Here $\bar{x}_t(x,p)$ is the classical one-dimensional trajectory of a  particle starting, at $t=-T$, with momentum $p$ at the position $x-p T/m$. It evolves in the effectively time-dependent potential $U_t(x)=V \left( x , t p_z/m \right)$. The associated momentum is $\bar{p}_t(x,p)$.
It is easy to see that the limits in (\ref{eqn:semiclassamp}), (\ref{eqn:semiclassphase}) exist for sufficiently short-ranged potentials. 
In particular, if the scattering potential has a finite extension the limit is reached already at finite times $T$, once the longitudinal distance $2T v_z$ is larger than the size of the interaction region. 

The reduced scattering factor (\ref{eqn:semiclassresult}) 
constitutes a significant improvement over the eikonal approximation used so far, as will be demonstrated in the next section. Compared to the elementary eikonal approximation (\ref{eqn:simpleeiko}) and to the Glauber eikonal result (\ref{eqn:glauber}), this expression not only provides the consistent incorporation of the deflection into the phase, but it also introduces an amplitude modification of the scattered wave. 

\section{Numerical Analysis} \label{sec:numerics}

We proceed to analyze the numerical performance of the semiclassical scattering factor (\ref{eqn:semiclassresult}) as
compared to the Glauber eikonal approximation (\ref{eqn:glauber}) and to the elementary eikonal approximation used so far in evaluations of the Talbot-Lau effect. We will see that the elementary eikonal approximation was appropriate in the molecular matter wave experiments performed to date \cite{Brezger2002a,Gerlich2007a}. At the same time, both the Glauber and the semiclassical approximation significantly improve the treatment of laser gratings if the particles have 
larger polarizabilities or smaller velocities (as required if their mass ins increased).

According to the general theory from Sect.~\ref{sec:generalTL} the interference pattern (\ref{eqn:interferencepattern2}) is determined by the generalized, momentum dependent grating coefficients (\ref{eqn:gratingcoeff2}) via the scattering factor (\ref{eqn:Sfactor_1D}). 
In order to assess the validity
of the different approximations it is therefore pertinent to evaluate
the scattering factors directly and compare them to a numerical implementation of the exact propagation of a plane wave through the grating interaction
region.

\subsection{The scattering factor}

The exact transverse scattering factor (\ref{eqn:Sfactor_1D}) for the grating can be obtained by a numerical evaluation of the limit
\begin{equation}
	\phi \left(x,p \right) = \lim_{T\to\infty} \frac{\la x | \oU_{-T}^{(0)} \oU_{2T} \oU_{-T}^{(0)} |p \ra}{\la x |p \ra}.
\end{equation}
In practice, this is done by computing the propagator matrix elements by means of a split operator technique \cite{Feit1982a,Chambers2000a}, making sure that the propagation time $T$ is sufficiently large (much larger than the grating passage time) so that the result is converged. In the following examples we use periodic boundary conditions for the position coordinate $x$ and a fixed transverse momentum $p$ not larger than the typical beam spread $10^{-3} p_z$.

\subsubsection{Material grating}

\begin{figure}
\includegraphics[width=8cm]{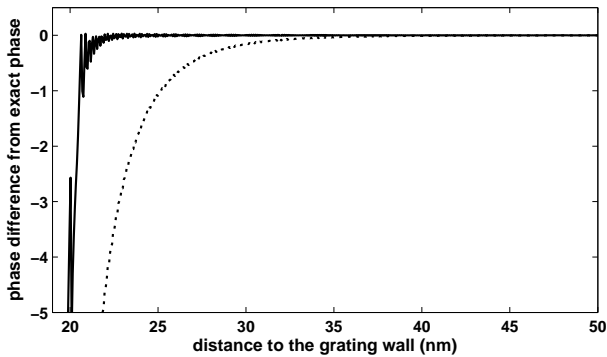}
\caption{\label{fig:scatt_mat} Phase difference of the semiclassical (solid line) and the eikonal approximation (dotted line) to the exact result, plotted versus the distance to one grating slit wall. We use grating and particle parameters as in Fig.~\ref{fig:lambda_mat} (corresponding to the fullerene experiment \cite{Brezger2002a}).}
\end{figure}

We first take the grating to be of material type with a retarded Casimir-Polder interaction potential. Figure~\ref{fig:scatt_mat} shows the phase \emph{differences} between the approximate scattering factors and the exact scattering factor as a function of the distance to the wall. The calculations were done for a de Broglie wavelength of $\lambda = 4\,$pm. Moreover, we choose an orthogonally incident plane wave, $p=0$, so that the elementary and the Glauber eikonal approximations coincide. They are given by the dotted line, while the solid curve corresponds to the semiclassical approximation (\ref{eqn:semiclassresult}).

As one observes in Figure~\ref{fig:scatt_mat}, the divergence of the wall potentials invalidates the approximations for the scattering factor close to the walls. However, the contributions from this small vicinity of the slit walls do not affect the interference visibility appreciably, since it corresponds only to a small fraction of the semiclassical trajectories.
This suggests that a reasonable cut-off criterion is given by the critical distance $x_c$ to the wall, where a classical beam particle would hit the wall within the grating passage time $t=b/v_z$. Disregarding the initial transverse momentum $p$ and the potential of the opposite wall, the time for a particle starting at the distance $x_0$ to hit the wall is given by
\begin{equation}
	T = \sqrt{\frac{m}{2}}  \int_0^{x_0} \diff x \, \frac{1}{\sqrt{- V \left(x \right)}} .
\end{equation}
For a wall potential $V(x) = -C_4 x^{-4}$ this leads to the critical distance
\begin{equation}
	x_c = \left( \frac{18m C_4 b^2}{p_z^2} \right)^{1/6}.
\end{equation}
The parameters used for Fig.~\ref{fig:scatt_mat} yield  $x_c=21\,$nm, and this value indeed corresponds to the position 
where the semiclassical and the exact phase deviate by about $2\pi$.
The fact that the eikonal approximation is virtually identical to the exact phase factor for most of the slit width explains why the eikonal approximation is well justified with thin material gratings. 
In fact, our numerical results indicate that in this case the eikonal approximation remains valid even for
particles with a stronger particle-wall interaction, breaking down only in a regime where the interference visibility is already strongly diminished by the interaction effect. We therefore focus on the KDTLI setup in the following. 

As a final point, we note that the opening width of the slits is effectively reduced by twice the critical distance $x_c$. For large and slow particles it may be necessary to take this into account even at the first and at the third grating, since the fringe visibility depends quite sensitively on the corresponding effective open fractions.

\subsubsection{Laser grating}

\begin{figure*}
\includegraphics[width=8cm]{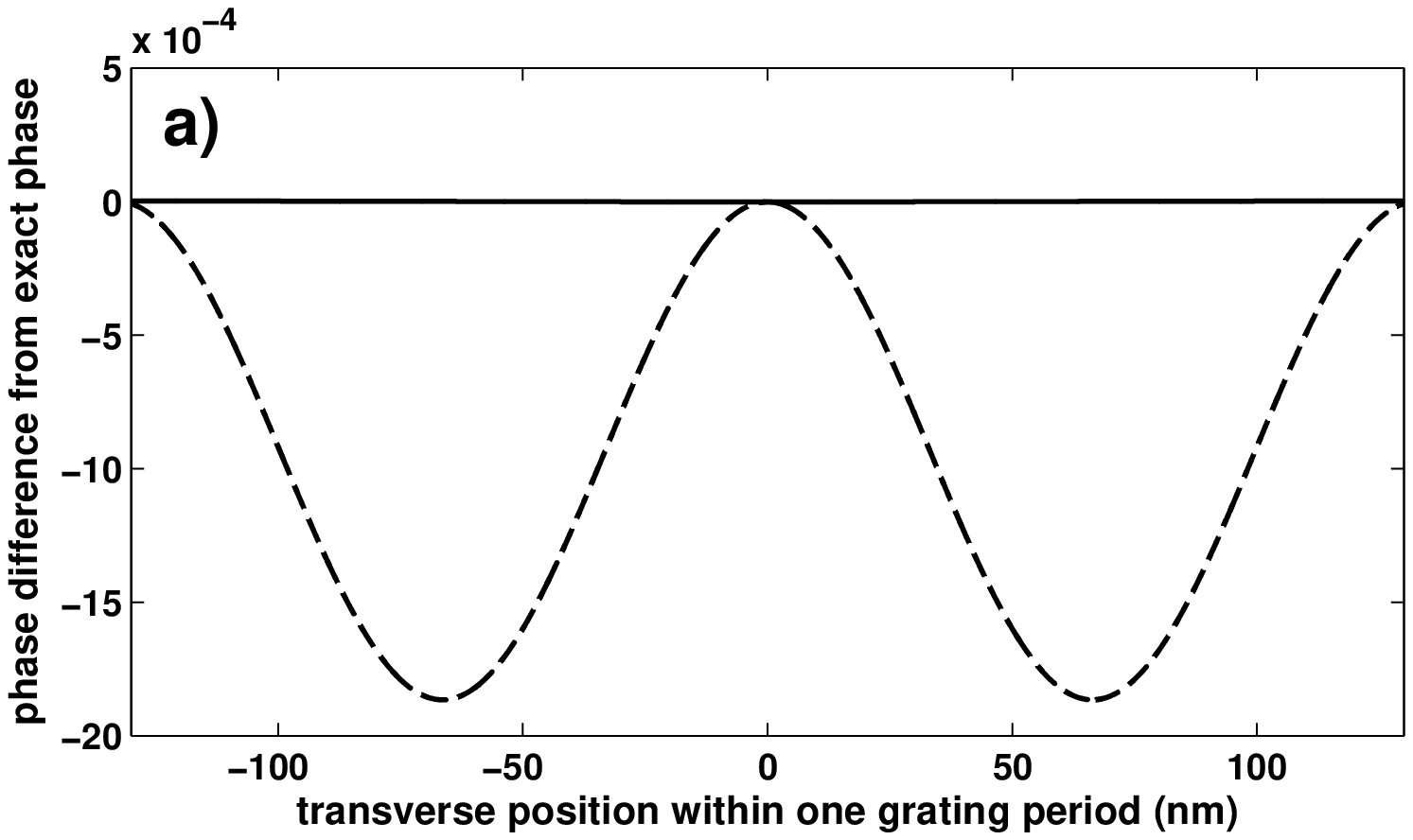} \hspace{5mm} \includegraphics[width=8cm]{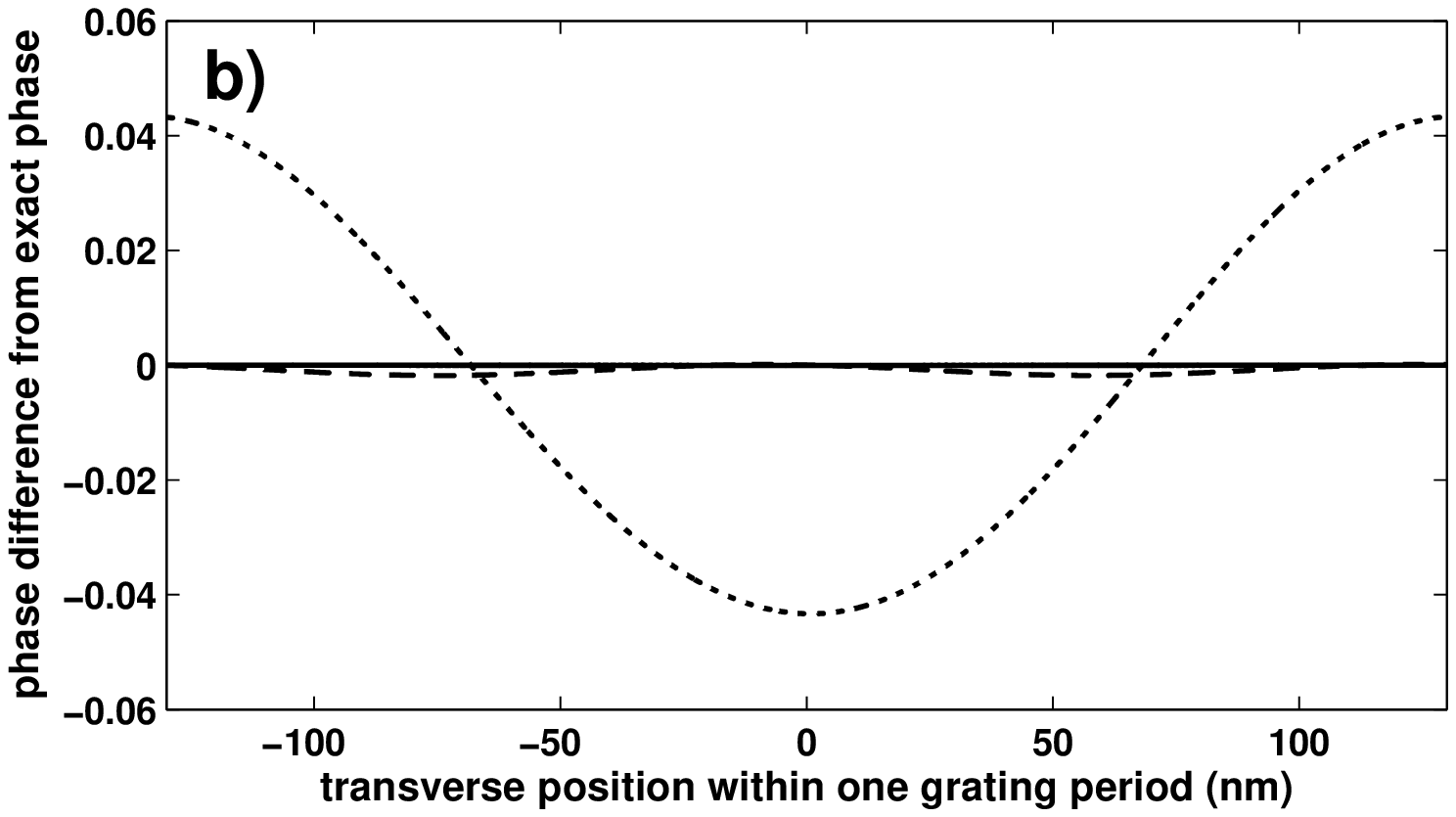}
\caption{\label{fig:scatt_laser1} Phase difference with respect to the 
exact scattering factor over one period $d=266\,$nm of the laser grating 
(parameters as in Fig.~\ref{fig:lambda_laser}, with $\lambda=3\,$pm, corresponding to the experiment \cite{Gerlich2007a}). Panel (a) shows the results for an incoming plane wave with vanishing transverse momentum, $p=0$, while  (b) corresponds to $p=10^{-3}p_z$. The solid line gives the
error of the semiclassical approximation, 
which is not resolved on the scale given by the errors of the 
Glauber approximation (dashed line) and the elementary eikonal approximation (dotted line). The latter are indistinguishable in (a); notice the different scales of the $y$-axes.}
\end{figure*}

Replacing the diffraction grating by a standing laser wave leads to the smooth and bounded interaction potential (\ref{eqn:laserpot}). 
For the numerical evaluation of the scattering factor, shown in Fig.~\ref{fig:scatt_laser1}, we choose the same parameters as for Fig.~\ref{fig:lambda_laser}
(motivated by the experiment \cite{Gerlich2007a}) and choose $\lambda=3 \,$pm. The longitudinal laser waist $w_z = 20\,\mu$m is much larger than the grating period $d=266 \,$nm. It follows that non-zero transverse momenta $p\neq 0$ now have to be considered separately since the free transverse motion over the distance $ w_z p/p_z$ must not be neglected. 

Figure~\ref{fig:scatt_laser1} shows how the phases of the approximate scattering factors deviate from the exact phase. Panel (a) corresponds to a perpendicular incidence of the incoming plane wave, $p=0$, while (b) is evaluated for $p = 10^{-3} p_z$, as found in a beam spread of $1\,$mrad 
The Glauber and the elementary eikonal approximation coincide in the case (a) of perpendicular incidence, and they deviate from the exact phase by about 1\,mrad. The error of the semiclassical phase is smaller by three orders of magnitude, and it is not resolved in the plot. On the other hand, in the case (b) of a non-zero transverse momentum the elementary eikonal approximation deviates substantially from the exact result, while the error for the Glauber approximation remains on the order of $10^{-3}$ and the semiclassical one on the order of $10^{-6}$ (not resolved in the plot). 

As for the corresponding amplitude of the incident plane waves, the exact calculation  yields deviations from the incident amplitude $1$ on the order of $10^{-6}$ in case (a) and deviations on the order of $10^{-4}$ in case (b), respectively (not shown). While the eikonal approximations cannot account for this effect, the semiclassical amplitude (\ref{eqn:semiclassamp}) reproduces the exact result with an error
of less than $10^{-8}$, and $10^{-6}$, respectively.

The semiclassical expression of the scattering factor is thus demonstrated to be superior by orders of magnitude compared to the eikonal approximations.
While the Glauber eikonal approximation already improves the elementary eikonal approximation significantly, it still does not take the amplitude modification into account. However, the overall corrections to the eikonal approximations are so small in the present parameter regime that the Talbot-Lau interference contrast is hardly affected, as demonstrated below. The elementary eikonal approximation thus remains valid for the considered experiment \cite{Gerlich2007a}.

\begin{figure*}
\includegraphics[width=8cm]{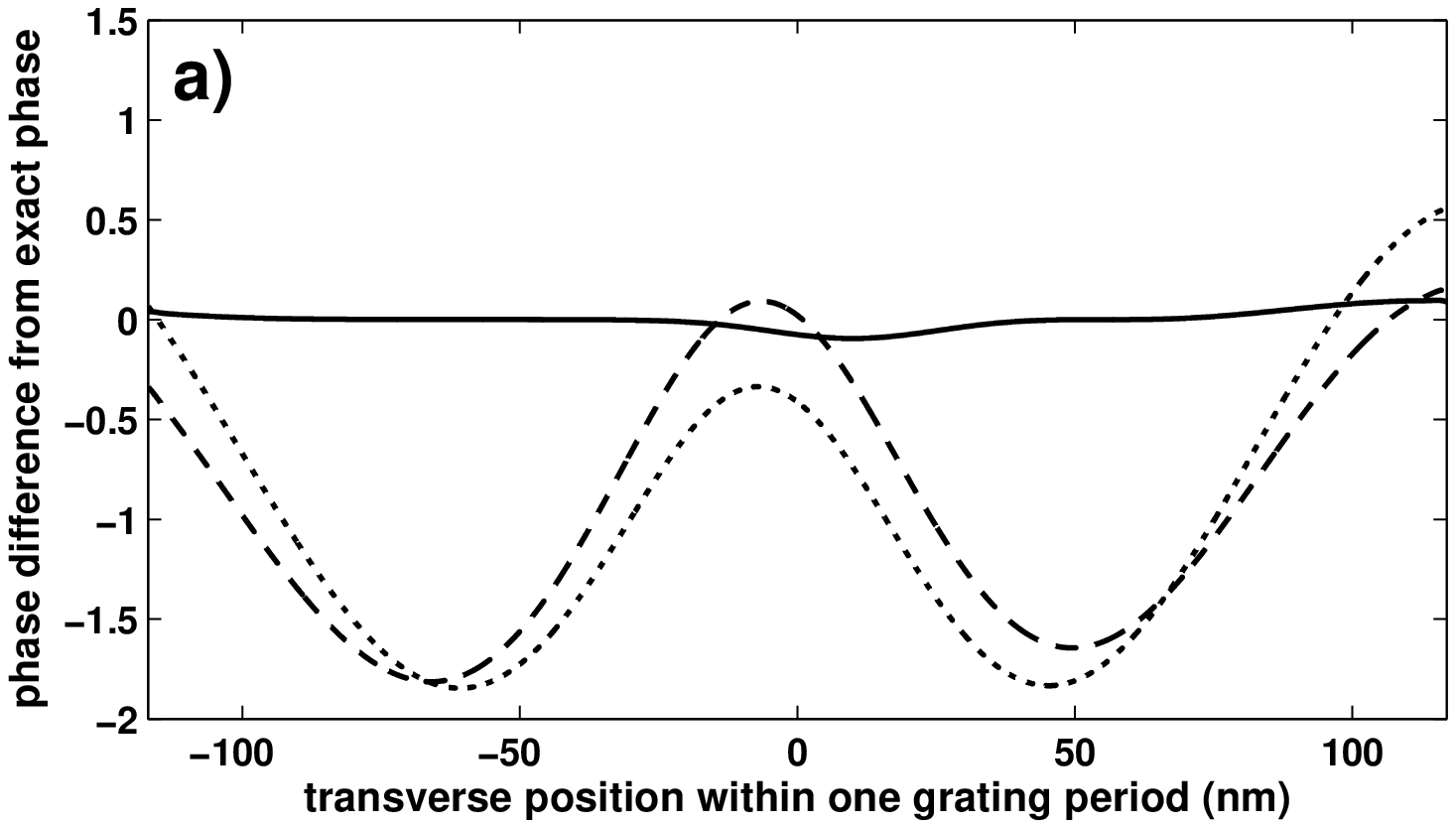} \hspace{5mm} \includegraphics[width=8cm]{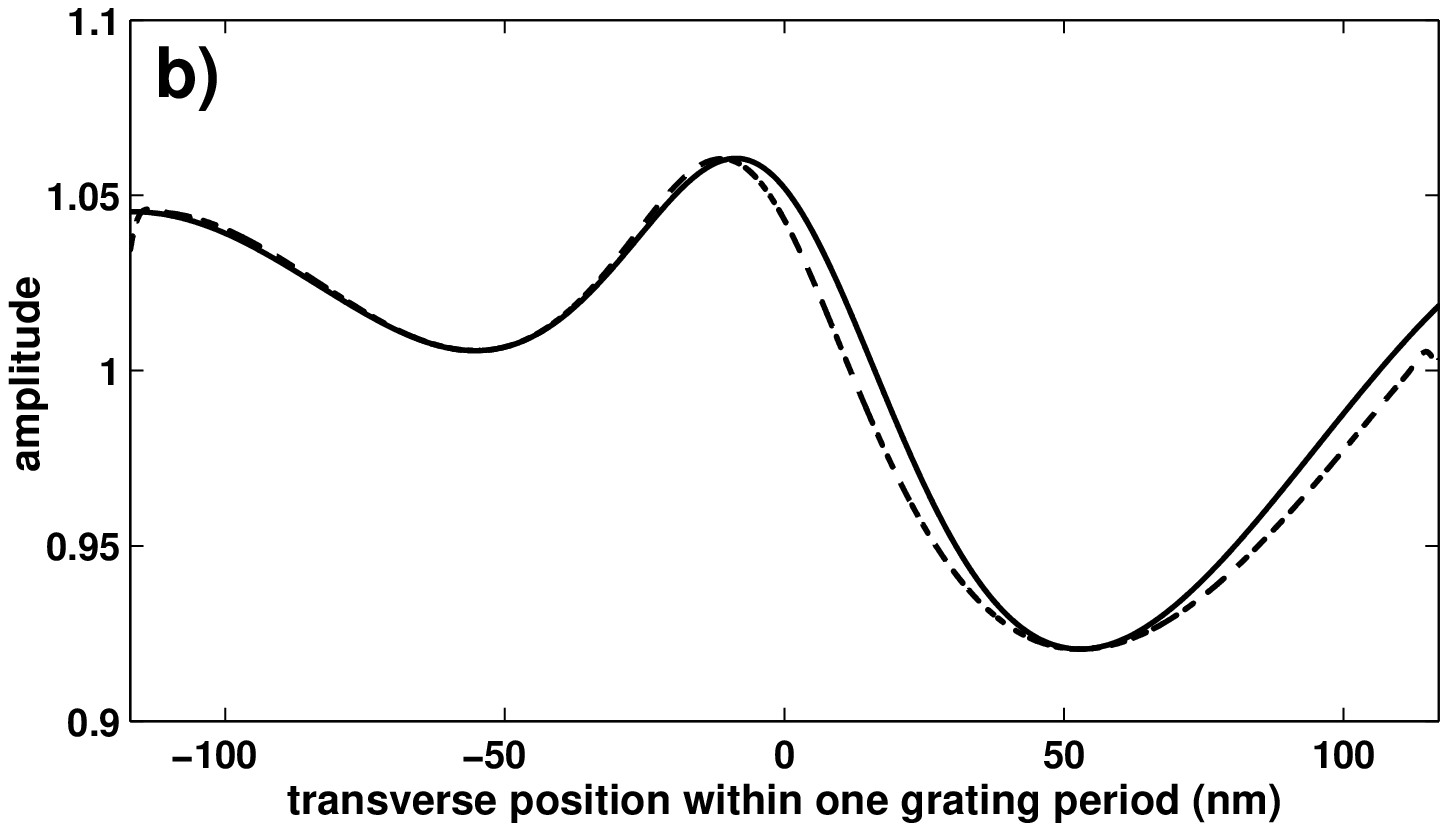}
\caption{\label{fig:scatt_laser2} (a) Phase differences with respect to the exact calculation like in Fig.~\ref{fig:scatt_laser1}, but for an increased de Broglie wavelength $\lambda=30\,$pm and $p=10^{-3}p_z$. The elementary eikonal approximation (dotted line) and the Glauber eikonal approximation (dashed line) now differ markedly from the exact phase result, while the semiclassical approximation (solid line) deviates by less that 100\,mrad.
(b) Corresponding amplitude modification 
as obtained from the semiclassical approximation (solid line) and the exact calculation (dashed line).}
\end{figure*}

The situation changes distinctively if the de Broglie wavelength is increased by a factor of ten, to $\lambda = 30\,$pm. The resulting phase difference and amplitude plots are shown in Fig.~\ref{fig:scatt_laser2} for the case of a transverse momentum $p=10^{-3}p_z$. Now both the Glauber and the elementary eikonal phase strongly differ from the exact result, as demonstrated by the dashed and the dotted curves in Fig.~\ref{fig:scatt_laser2}(a). At the same time, the semiclassical result (solid line) deviates by less than $100\,$mrad from the exact phase, and also the corresponding amplitude, seen in Fig.~\ref{fig:scatt_laser2}(b), faithfully approximates the exact one.

The semiclassical expression (\ref{eqn:semiclassresult}) starts to fail only if we decrease the beam velocity to such an extent that the trajectories get strongly deflected during the increased passage time. This is expected since the derivation assumes the corrections due to deflection to be small. Equation (\ref{eqn:semiclassresult}) thus extends the eikonal approximation to the smaller beam velocities required for more massive particles, but it does not cover the whole semiclassical wavelength regime. 

\subsection{The Talbot-Lau visibility}

\begin{figure}
\includegraphics[width=8cm]{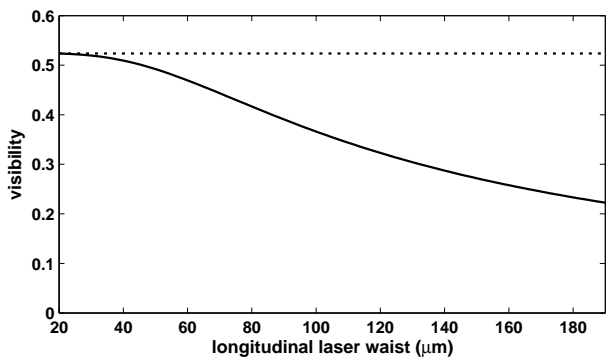}
\caption{\label{fig:vis_wz} Interference visibility as a function of the longitudinal waist $w_z$ of the laser grating (for $\lambda = 3\,$pm, starting from the experimental value $w_z=20\, \mu$m  \cite{Gerlich2007a}). The dotted line gives the result of the elementary eikonal approximation (\ref{eqn:simpleeiko}), while the solid curve represents the semiclassical result (\ref{eqn:semiclassresult}).}
\end{figure}

\begin{figure}
\includegraphics[width=8cm]{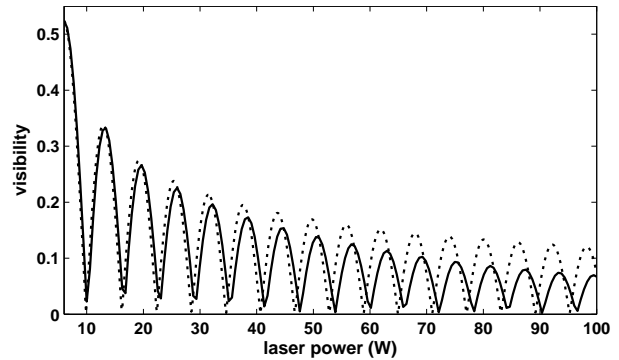}
\caption{\label{fig:vis_P} Fringe visibility  as a function of the laser power $P_L$, starting from the experimental value $P_L=6\,$W \cite{Gerlich2007a}. For considerably larger powers, or, equivalently, for larger polarizabilities of the particles, the elementary eikonal approximation (dotted line) deviates markedly from the semiclassical result  (\ref{eqn:semiclassresult}).}
\end{figure}

\begin{figure}
\includegraphics[width=8cm]{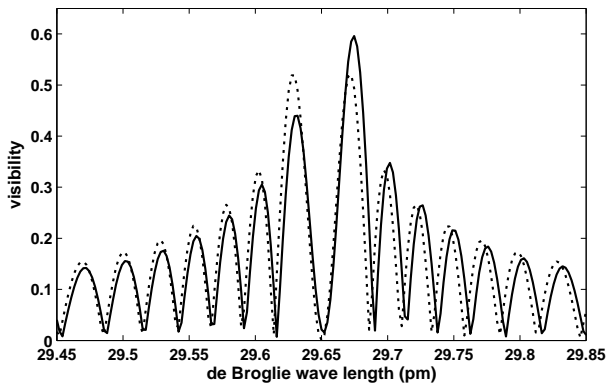}
\caption{\label{fig:vis_lambda} Talbot-Lau interference visibilities for the laser grating setup \cite{Gerlich2007a}
with C$_{70}$ fullerenes. We use increased de Broglie wavelengths 
corresponding to velocities around $v_z=16$\,m/s \cite{Deachapunya2008a}.
The semiclassical approximation (solid line) yields values which are clearly distinguishable from the 
Glauber eikonal approximation (dotted line). The elementary eikonal approximation differs 
more strongly.}
\end{figure}

We can now discuss how the improved treatment of the grating interaction effect affects the Talbot-Lau interference visibility. We  focus again on the laser grating setup demonstrated in \cite{Gerlich2007a}. 
In the Figures \ref{fig:vis_wz}--\ref{fig:vis_lambda}
the eikonal results are obtained by calculating the visibility (\ref{eqn:Vexp}) 
by means of the Talbot-Lau coefficients (\ref{eqn:Beiklaser}). The semiclassical calculation implements the generalized formula for the interference pattern (\ref{eqn:interferencepattern2}), where the semiclassical scattering factor (\ref{eqn:semiclassresult}) enters by means of the generalized grating coefficients (\ref{eqn:gratingcoeff2}). Note that, unlike in the elementary eikonal approximation, it is now essential to incorporate the angular distribution $D \left( p / p_z \right)$ into the calculation.
It is set here to be a Gaussian $D \left( p / p_z \right) \propto \exp \left( -(p/p_z)^2 / 2 \alpha^2 \right) $ with a realistic width of $\alpha = 1\,$mrad.

One immediate consequence of the dependence of the semiclassical and the Glauber approximations on the transverse momentum is demonstrated in Fig.~\ref{fig:vis_wz}, where the interference visibility is plotted versus the longitudinal laser waist $w_z$, starting from the experimental value $w_z = 20 \, \mu$m. The elementary eikonal approximation (dotted line) is independent of $w_z$ due to the longitudinal integration in (\ref{eqn:simpleeiko}). The semiclassical result (solid line), which takes into account the transverse motion through the laser field, decreases with growing $w_z$. While the difference between the approximations is negligible at the experimental value, it becomes significant for a larger laser focus. The Glauber approximation reproduces the semiclassical result up to a precision of $10^{-4}$ and is therefore indistinguishable from the solid line Fig.~\ref{fig:vis_wz}. This implies that the visibility loss with growing waist is due to the free transverse motion through the laser grating, rather than due to a considerable deflection of the trajectories.

A similar result is presented in Fig.~\ref{fig:vis_P}, where we increase the laser power starting from the experimental value $P_L=6\,$W. This is equivalent to increasing the polarizability of the particles, see (\ref{eqn:laserpot}). One observes that the semiclassical result (solid curve) decreases more rapidly than the eikonal visibility (dotted curve) as the strength of the phase grating is increased. The Glauber approximation again matches the semiclassical curve.

Finally, in Fig.~\ref{fig:vis_lambda} we present the interference visibility as a function of the de Broglie wavelength $\lambda$ around $30\,$pm, corresponding to a tenfold smaller beam velocity than in the experiment. In this case one observes at some wavelengths a considerable difference between the 
Glauber eikonal approximation (dotted line) and the semiclassical approximation (solid line). 

Note that the sharp dips, where the visibility drops to zero in Fig.~\ref{fig:vis_lambda} and Fig.~\ref{fig:vis_P}, indicate a shift of the interference pattern by half its period. One has to take this into account when averaging the visibility over the velocity distribution of the particle beam. For example, if this distribution was so broad that it amounts to averaging over two subsequent peaks in Fig.~\ref{fig:vis_lambda} it would lead to almost zero visibility.

We have seen that in general the validity of the eikonal approximation depends on the grating interaction strength, the passage time, the longitudinal velocity, and the distribution of the transverse momenta in the particle beam. For the specific experiment \cite{Gerlich2007a}  the elementary eikonal approximation breaks down by either decreasing the longitudinal velocity by a factor of $10$, or similarly by increasing the laser power $P_L$, the longitudinal waist $w_z$, or the particle polarizability by an order of magnitude. The latter are effects mainly due to the transverse motion of the beam particles, which is not taken into account by the elementary eikonal approximation. Here the Glauber eikonal approximation (\ref{eqn:glauber}), i.e., the semiclassical scattering factor without deflection, would be already sufficient, at least in the experimentally accessible regime. This is no longer the case 
if one increases the wavelength, since the deflection effect is more sensitive to the wavelength than to the grating parameters, limiting the validity of both the elementary and the Glauber eikonal approximation. 

\section{Conclusions} \label{sec:conclusions}

We presented a general theory of the coherent Talbot-Lau interference effect.  
It allows to incorporate the interaction between particle and grating structure--a dominant effect in near field interference--at various degrees of approximation. Our treatment shows that it is necessary to account for detailed beam characteristics, such as the angular distribution, whenever one is required to go beyond the elementary eikonal approximation, or if one wants to quantify the experimental adjustment requirements.

Using the phase space formulation of quantum mechanics, we identify the appropriate generalization of the Talbot-Lau coefficients.
They serve to incorporate the most general coherent grating transformation and to describe the various near field interference effects in a transparent fashion. The general effect of the passage through a grating can thus be formulated in terms of scattering theory, providing a starting point for the numerically exact evaluation of the interference pattern. 

Moreover, the semiclassical approximation of the S-matrix  yields a systematic and non-perturbative improvement over the elementary eikonal approximation. 
An additional high-energy approximation of the semiclassical trajectories then yields the Glauber eikonal approximation and the semiclassical deflection approximation as systematic corrections to the standard treatment.
A comparison with the numerically exact calculation verifies the high quality of the semiclassical deflection approximation. It suggests that a Kapitza-Dirac Talbot-Lau interferometer, where the center grating is replaced by a standing light wave, will be able to demonstrate the wave nature even of particles which are so large that the eikonal approximation is no longer valid. 

\begin{acknowledgments}
We thank M. Arndt and H. Ulbricht for helpful discussions.
This work was supported by the FWF doctoral program `Complex Quantum Systems' (W1210) and by the DFG Emmy Noether program.
\end{acknowledgments}

\appendix*

\section{Classical description} \label{app:class}

If the Talbot-Lau experiment is to prove the quantum nature
of particles one clearly needs to be able to distinguish between the quantum
interference effect and the moir\'{e}-type shadow effect that may occur with classical particles. A formulation is therefore required that yields the classical shadow contrast by using the same assumptions and approximations as in the quantum case. We present this classical theory in the following by assuming a material grating of thickness $b$ with a transverse interaction potential $V(x)$ of the
grating slit walls. The case of an explicitly $z$-dependent
interaction potential $V \left(x,z \right)$, such as a laser grating \cite{Gerlich2007a}, can be treated in the
one-dimensional model by an effectively time-dependent potential
$\widetilde{V} \left(x,t \right) = V \left(x, p_z t /m \right)$, where
the longitudinal motion provides the time coordinate for a given
momentum $p_z$. 

The classical formulation is based on the phase space density rather than the Wigner function. The corresponding classical propagator through a diffraction
grating is given by the expression
\begin{eqnarray}
 K_{\text{cl}} \left(x,p;x_0,p_0 \right) &=& \left|t\left(x_0\right) \right|^2 \delta \left( x_0 - x_0^{\text{cl}} \left(x,p,\frac{mb}{p_z} \right) \right) \nonumber \\
&&\times \, \delta \left( p_0 - p_0^{\text{cl}} \left(x,p,\frac{mb}{p_z} \right) \right),
\label{eqn:klassprop}
\end{eqnarray} 
where the hard grating wall cutoff of the particle beam is taken into
account at the entrance into the grating. The phase space coordinates
$x_0^{\text{cl}},p_0^{\text{cl}}$ are the starting point of the
classical trajectory evolving to $x,p$ under the influence of the
interaction potential $V(x)$ within the grating passage time
$t=mb/p_z$. 

Since the free evolution of the phase space density
is given by the same transformation (\ref{eq:wt}) as in the case of the Wigner function,
one
ends up with the general classical shadow pattern, denoted by $f_3$
instead of $w_3$, after
substituting the classical propagator (\ref{eqn:klassprop}) into
the general Talbot-Lau calculation from Sect.~\ref{sec:generalTL}.
\begin{widetext}%
\begin{eqnarray}
 f_3 \left( x \right) &=& \frac{1}{G} \int \frac{\diff p}{p_z} \, D \left( \frac{p_0^{\text{cl}} \left( x- \frac{p}{p_z}\eta L,p,\frac{mb}{p_z}\right)}{p_z}\right) \left|t \left( x_0^{\text{cl}} \left( x- \frac{p}{p_z}\eta L,p,\frac{mb}{p_z}\right)\right) \right|^2 \nonumber \\
&& \times \, \left|t_1 \left( x_0^{\text{cl}} \left( x- \frac{p}{p_z}\eta L,p,\frac{mb}{p_z}\right) - \frac{p_0^{\text{cl}} \left( x- \frac{p}{p_z}\eta L,p,\frac{mb}{p_z}\right)}{p_z} L\right) \right|^2
\label{eqn:shadow1}
\end{eqnarray} 
\end{widetext}
One can numerically implement this formula directly, rather than performing a Fourier decomposition with respect to the argument $x-p \eta L /p_z$ of the trajectory terms. 

The ideal moir\'{e} shadow pattern is obtained if one disregards both the interaction potential and the grating thickness by setting $x_0^{\text{cl}} \left(x,p,t\right)=x$ and
$p_0^{\text{cl}} \left(x,p,t\right)=p$. 
If a particle-wall interaction is present, the classical analogue to the eikonal approximation (\ref{eqn:simpleeiko}) is to approximate the deflection of the trajectory due to the grating by an instantaneous momentum kick \cite{Hornberger2004a},
\begin{eqnarray}
 x_0^{\text{cl}} \left( x, p, \frac{m b}{p_z} \right) &=& x \nonumber \\
 p_0^{\text{cl}} \left( x, p, \frac{m b}{p_z} \right) &=& p + \frac{m b}{p_z} V'(x).
\end{eqnarray}
Putting this into the classical formula (\ref{eqn:shadow1}), performing all the Fourier decompositions, and focusing on a particular $r:s$ Talbot-Lau resonance, as  done in the eikonal quantum case (\ref{eqn:w3_eik2}), one obtains the classical shadow pattern in eikonal approximation,
\begin{eqnarray}
 f_3 \left( x \right) &=& \frac{1}{G} \sum_{\ell=-\infty}^{\infty} A_{s \ell}^{*} \exp \left(2 \pi i \ell\frac{ x}{d_3} \right) \nonumber \\
&&\times \, \sum_{k=-\infty}^{\infty} B_{r\ell-k} (0) c_k \left( s \ell \frac{d}{d_1}\frac{L}{L_\textrm{T}} \right). \label{eqn:f3_eik}
\end{eqnarray} 
The classical momentum kick coefficients read as
\begin{eqnarray}
 c_n \left( \xi \right) &=& \frac{1}{d} \int_{-d/2}^{d/2} \diff x \, e^{-2 \pi i n x / d} \exp \left( i \xi \frac{m b V' (x) / p_z}{\hbar / d} \right) .
\nonumber\\&&
\end{eqnarray} 
If the second grating is implemented by a standing laser beam  the
classical calculation  yields an analytical expression for the
Talbot-Lau coefficients $B_m^{\text{cl}} \left( \xi \right) = \sum_k
B_{m-k} \left(0 \right) c_k \left( \xi \right)$. They are related to the
quantum expression (\ref{eqn:Beiklaser}) by replacing the sine function in the argument with its  linear expansion,
\begin{equation}
	B_m^{\text{cl}} \left(\xi \right) = J_m \left( - \frac{4 M P_L \alpha_{\omega} }{\sqrt{2\pi} \hbar c \varepsilon_0 w_y p_z}  \pi \xi \right).
	\label{eqn:Beiklasercl}
\end{equation}
Since the argument of the $B_m$ is proportional to the de Broglie
wavelength $\lambda$ in the Talbot-Lau interference effect, this means
that the quantum
interference and the classical shadow effect become indistinguishable in the naive classical limit of a vanishing wavelength, $\lambda \rightarrow 0$.

\end{document}